%% file: manuscript_v11.tex
\documentclass[10pt, final, twocolumn, twoside, romanappendices]{IEEEtran}
\usepackage{amsmath,amssymb,bbm}
\usepackage[level=3]{LatexInclusion/wgroup_message}  % set level to 0, to hide all notes
\input{LatexInclusion/Wgroup-Teaching-Preamble}

\usepackage[bookmarks,colorlinks]{hyperref} % create hyperlinks
\hypersetup{colorlinks,citecolor= red,filecolor= blue,linkcolor= blue,urlcolor=blue}
\usepackage{acronym}  % make an acronym
\usepackage{color}  % make colors
\usepackage[usenames,dvipsnames]{xcolor}
\usepackage{amsthm}
\usepackage{amsfonts}
\usepackage{dsfont}
\usepackage{mathrsfs}
\usepackage{pifont}
\usepackage{amssymb}
\usepackage{verbatim}
\usepackage{upgreek}
\usepackage[final]{graphicx}
\usepackage{caption}
\usepackage{algorithmic}
\usepackage{algorithm}
\usepackage{epsfig}
\usepackage{eucal}
\usepackage{cite}
\usepackage{subfigure}
\usepackage{stmaryrd}

\UseRawInputEncoding

\input{LatexInclusion/defmetric.tex}
\input{LatexInclusion/acronym}

\usepackage[yyyymmdd,hhmmss]{datetime} % gives time of the compiling
\newdateformat{monthyeardate}{\monthname[\THEMONTH] \THEDAY, \THEYEAR} % specifies the format of the date

\newcommand{\versionnumber}{07}

\newcommand{\versionDT}{\textcolor{BLUE}{\textsc{Version: V\versionnumber -- \monthyeardate\today\\
														Time: \currenttime}}}													

\newcommand{\paperTitle}{Anomaly Detection Algorithms for Location Security in 5G Scenarios}
\newcommand{\paperTitleMarkboth}{Anomaly Detection Algorithms for Location Security in 5G Scenarios}

\begin{document}

{
%%%%% UNCOMMENT THIS LINE IF YOU HAVE 2-column version %%%%%%
\twocolumn
%%%%% UNCOMMENT THIS LINE IF YOU HAVE 2-column version %%%%%%

%---------------
% Release notes:
%   Jan 25, 2012 -- Ae added an example of how to use acronym package
%   Jan  5, 2012 -- Ae updated the statement above the sponsor (in title), 
%                   made the cover sheet 1-column (in section cover sheet), and
%                   added a command "\paperTitle" (for section cover sheet and title).
%   June 2, 2011 -- Ae updated the questions in "WGroup Research Paper Catechism"
%                   (using the questions from Prof. Win and Andrea Conti)
%---------------

%---------------------------------------------------------------------------%
%                     title, title footnote, header                         %
%---------------------------------------------------------------------------%

% paper title
\title{\paperTitle}

% Uncomment this line, if it's an invited paper
% \IEEEspecialpapernotice{(Invited Paper)}

% author names, IEEE memberships, corresponding address, title footnote %
\author{
%%%%%%%% uncomment this section for a 2-column format %%%%%%%
%%%%%%%% [begin] %%%%%%%%
	\vspace{0.2cm}
%%%%%%%% [end] %%%%%%%%	
    Stefania Bartoletti,~\IEEEmembership{Member,~IEEE},
	Ivan Palam\`a,    
    Danilo Orlando,~\IEEEmembership{Senior Member,~IEEE},
    Giuseppe Bianchi,
    Nicola Blefari Melazzi,~\IEEEmembership{Senior Member,~IEEE}\\
	\vspace{0.4cm}
%	\callout
	\vspace{-0.2cm}  
    %~~~~~~~~~~~~~~~~~~~~~~~~~~~%
    %       Mailing Address     %
    %~~~~~~~~~~~~~~~~~~~~~~~~~~~%
%%%%%%%% comment this section for a 2-column format %%%%%%%
%%%%%%%% [begin] %%%%%%%%
%    \\\vspace{1.5cm} % vertical space of length 2 cm
%    \underline{Corresponding Address:}\\
%    Watcharapan~Suwansantisuk\\
%    Laboratory for Information and Decision Systems (LIDS)\\
%    Massachusetts Institute of Technology (MIT)\\
%    77 Massachusetts Avenue, Room 32-D674A\\
%    Cambridge, MA 02139 USA\\
%    \bigskip
%    Tel.: (617) 324-1548\\
%    e-mail: {\tt wsk@mit.edu}
%%%%%%%% [end] %%%%%%%%
    %~~~~~~~~~~~~~~~~~~~~~~~~~~~%
    %      Title footnote       %
    %~~~~~~~~~~~~~~~~~~~~~~~~~~~%    
    \thanks{This work was supported by the European Union's Horizon
2020 research and innovation programme under Grant
no. 871249.
%		the National Science Foundation under Grant CCF-1525705,
%%  		the Office of Naval Research under Grant N00014-11-1-0397,
%		the Office of Naval Research under Grant N00014-16-1-2141,
%%		the Defense University Research Instrumentation Program under Grant N00014-08-1-0826,
%		the Defense University Research Instrumentation Program under Grant N00014-17-1-2379,
%		the Italian MIUR project GRETA under Grant 2010WHY5PR,
%		the Copernicus Fellowship,
%		and
%		the MIT Institute for Soldier Nanotechnologies. 
%   	%---- This disclosure should be is accurate before submission -----
%	The material in this paper was presented, in part,
%		at the IEEE Global Telecommunications Conference, Atlanta, GA, December 2013, and 
%		at the IEEE International Conference on Communications, Sydney, Australia, June 2014.
    }
    \thanks{Stefania Bartoletti is with the National Research Council of Italy, CNR-IEIIT and CNIT, viale Risorgimento 2, 40136 Bologna, Italy. E-mail: {\tt stefania.bartoletti@cnr.it}.}
    \thanks{Danilo Orlando is with the Engineering Faculty of Universit\`a degli Studi ``Niccol\`o Cusano'', 
via Don Carlo Gnocchi 3, 00166 Roma, Italy. E-mail: {\tt danilo.orlando@unicusano.it}.}
    \thanks{Ivan Palam\`a, Giuseppe Bianchi, and Nicola Blefari Melazzi are with  Department of Electronic Engineering, Univ. of Roma Tor Vergata and CNIT, Via del Politecnico, 1 00133, Rome, Italy. Email: {\tt \{giuseppe.bianchi,blefari\}@uniroma2.it}}
    
%%%    \thanks{
%%%    
%%%%        
%%%	}
%	\thanks{
%	Color versions of one or more of the figures in this paper are available online at http://ieeexplore.ieee.org.
%	}
%	\thanks{
%	Digital Object Identifier 10.1109/TSP.2016.1234567
%	}	
	 
}

% make the title area
% Don't write page number 0 to the cover page.
\maketitle 

%\markboth{}{}
%\markboth{IEEE JOURNAL ON SELECTED AREAS IN COMMUNICATIONS, VOL.~26, NO.~1, JANUARY 2008}{}

%\markboth{PLEASE DO NOT DISTRIBUTE WITHOUT THE WRITTEN CONSENT OF THE AUTHORS \versionbox}
%		{Suwansantisuk, Chiani, and Win: \paperTitleMarkboth}
\markboth{PLEASE DO NOT DISTRIBUTE WITHOUT THE WRITTEN CONSENT OF THE AUTHORS}
		{Bartoletti, Palam\`a, Orlando, Bianchi, and Melazzi: \paperTitleMarkboth}

%%%%%%%% uncomment this section for a 2-column formt %%%%%%%
%%%%%%%% [begin] %%%%%%%%
%\thispagestyle{empty}
%\textcolor{blue}{\framebox{\textsf{Today: \today}}}

%\newpage
%%%%%%%% [end] %%%%%%%%

\setcounter{page}{1}

%---------------------------------------------------------------------------%
%                           abstract and key words                          %
%---------------------------------------------------------------------------%
\begin{abstract}
Location based services are expected to play a major role in future generation cellular networks, starting from the 
incoming 5G systems. At the same time, localization technologies may be severely affected by attackers capable to deploy 
low cost fake base stations and use them to alter localization signals. In this paper, we concretely focus on two classes of threats:
noise-like jammers, whose objective is to reduce the signal-to-noise ratio,
and spoofing/meaconing attacks, whose objective is to inject false or erroneous information into the receiver.
Then, we formulate the detection problems as binary hypothesis tests and solve them
resorting to the generalized likelihood ratio test design procedure as well as the Latent Variable Models, which
involves the expectation-maximization algorithm to estimate the unknown data distribution parameters. 
The proposed techniques can be applied to a large class of location data
regardless the subsumed network architecture. The performance analysis is conducted over simulated
data generated by using measurement models from the literature and highlights the effectiveness
of the proposed approaches in detecting the aforementioned classes of attacks.
\end{abstract}


\acresetall		% reset the acronyms 

%---------------------------------------------------------------------------%
%                                Introduction                               %
%---------------------------------------------------------------------------%
\section{Introduction}\label{sec:intro}
The incoming fifth generation mobile communication network (5G) will be capable of guaranteeing new reliable services, 
high data rates, low latency, and support for largely heterogeneous devices \cite{7839836}. Localization services, 
which in the past were mainly provided by non-cellular technologies (e.g., Global Navigation Satelline Systems and/or Global Positioning Systems) integrated within the cellular devices, are now conceived as first class citizens of the 5G architecture starting from Release 16  \cite{3GPP:TR:38.855:V16.0.0,3GPP:TS:37.355:V16.2.0,3GPP:TS:38.305:V16.2.0}, and will play a  crucial role in several scenarios envisioned for 5G, including self-driving cars, unmanned aerial vehicles, smart logistics, 
emergency services, and many more \cite{8579209,8795429,9120719}. 
At the same time, the ability to exploit location signals emitted by 5G base stations comes with a bleak side: many literature works 
\cite{7452266,7343507,6618585,rupprecht-19-layer-two,mjlsnes2017easy,rao2017lte,borgaonkar2019new,Shaik2019NewVulnerabilities,shaik2015practical,palama-imsi-catcher,jover2016lte,hussain-LTEinspector,Yu-LTEPhoneNumberCatcher} show how easy 
is for a tech savvy opponent to build ultra-low cost Jammers or even LTE/5G ``rogue'' base stations \cite{3GPP:TR:33.809:V17.0.0} 
capable to generate fake signals or interfere with legitimate ones.
Moreover, since in 5G a large number of stages interact towards the position estimation, there exists a plethora of points of the 
distributed communication system that can be object of malicious actions. 
Therefore, it emerges the strong need to equip the 5G network with suitable countermeasures aimed at preserving location data integrity.

Among the possible threats to location security, here we focus on intentional interference, which involves hostile platforms that 
target the \ac{UE} and/or \ac{AN} receivers in order 
to \cite{SmartJammingAttacksRev,3GPPlocSec,LiyAhmAbrGurYli:B17,SurveyJammSat}: (i) reduce the signal-to-noise ratio (\emph{noise-like jamming}); and (ii) inject false or erroneous information (\emph{spoofing/meaconing}).
(see also Figure \ref{fig:OperatingScenario}, where the ANs are denoted by gNB)
%\begin{itemize}
%\item 
%\item 
%\end{itemize} 

\begin{figure}[!t]
\centering
\includegraphics[width=0.7\columnwidth]{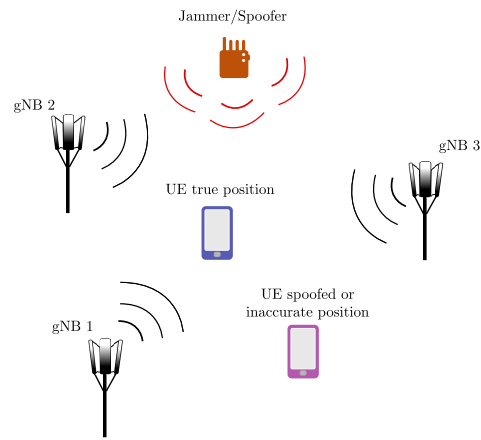}
\caption{Operating scenario for a cellular network under the attack of a jammer or spoofer which leads to inaccurate or counterfeit 
location estimates.}
\label{fig:OperatingScenario}
\end{figure}

The hostile platforms belonging to the first class of threats perform a denial of service attack by transmitting high-power interfering 
signals to disrupt the  functionalities of the receiver and are referred to as \acp{NLJ} 
\cite{MarMotAngFloFerPao:17,HenMakDomBobSanGao:14,SurveyJammSat}. 
In fact, the signals transmitted by NLJs blend into the thermal noise of the receiver leading to an increase of the noise power spectral 
density within the receiver bandwidth and, hence, to a decrease of the \ac{SNR}.
Remarkably, under the attack of NLJs, the estimation quality \ac{DOA}, \ac{TOA}, \ac{TDOA}, and \ac{RSS} measurements at 
a generic \ac{AN} degrades and leads to inaccurate \ac{UE} localization \cite{GroverJamming}. Most of the existing
detection and mitigation strategies to cope with NLJ attacks in wireless networks take place at physical and/or protocol layer \cite{MilitaryCommunications}.
%For instance, in , the authors investigate the effects of noise-like interference on the Physical Uplink 
%Control Channel in LTE and propose a mitigation strategy based upon the Radio Link Control protocol. 
Anti-jamming techniques in the context of cognitive radio networks are addressed in \cite{IEEEnetwork}, where several 
mitigation techniques are reviewed and a new anti-jamming protocol relying on probabilistic pairing and frequency tuning is proposed. 
Other examples of electronic counter-countermeasures are given in \cite{GameTheory1,GameTheory2} 
where an approach based on game theory is applied at the design stage. 
Mitigation solutions conceived at the physical layer can be found in \cite{TOPAL2020101029}, 
where the received signal is classified by means of a deep convolutional neural network 
fed by the signal features in the wavelet domain. Finally, in \cite{JammingMIMO02}, the authors leverage random matrix theory tools 
to conceive a multiple hypothesis test for jamming detection. To this end, they estimate the jammer 
subspace through the sample covariance matrix \cite{VanTrees4} and project the received data onto the user subspace in order
to mitigate the jamming components.

Platforms perpetrating the second mentioned class of attacks are aimed at deceiving the \ac{AN} receiver by injecting 
fake information into the latter. These deceptive
operations are commonly referred to as spoofing/meaconing \cite{LiyAhmAbrGurYli:B17,SmartJammingAttacksRev,SurveyJammSat}  
and intercept the positioning messages exchanged by two legitimate actors and 
suitably delay or modify them by synthesizing counterfeit information. Their main objective consists in preventing the victim systems 
from providing reliable position estimates. Deceptive attacks do not degrade 
the estimation quality of the location parameters but they are capable of changing the true parameter values.
%A breakthrough upgrade of 5G network with respect to 4G is the use of multiple antennas and, in particular, 
%of  which represent the key technology to increase the spectral efficiency 
%in the new wireless networks generation. 
In the context of massive \ac{MIMO} systems, a key feature of 5G networks, in \cite{JammingMIMO}, 
the \ac{GLRT} is applied to detect jamming signals contaminating user pilots. To this end, received raw 
data is first projected onto the subspace of the unused pilots in order to remove useful signal components and then 
the \ac{GLRT} is derived by estimating the unknown parameters through the maximum likelihood approach.
In \cite{JammingMIMO02}, a pilot spoofing attack detection algorithm in massive \ac{MIMO} systems is proposed, in
which the information of channel statistics is unknown. First, users send pilots to the base station,
then the latter transmits the conjugate of its received signal (which may contain spoofing
signal) back to users, where the final detection is made.

Summarizing, the above contributions show that intentional interference detection/mitigation in wireless networks 
can be accomplished at different layers according to the desired trade-off between computational requirements and effectiveness. 
In fact, it is clear that processing raw data (physical layer) might lead to more reliable techniques at the price of an increased complexity, 
whereas algorithms fed by high-level data (namely, the measurements provided by the network sensors) might be more efficient. 
In addition, in the latter case, since a 
preliminary processing is performed on data, the available information is limited and, more importantly, affected by 
the uncertainties introduced by the early stages \cite{ConMazBarLinWin:J19,MazConAllWin:J18,WinSheDai:J18}. Nevertheless, in
many situations raw data are not available given the specific architecture solutions and, hence, 
possible location security algorithms are limited to work with the measurements provided by the network infrastructure. 

It is important to stress here that the \ac{NR}-positioning work started with release 16 and  \cite{3GPP:TR:38.855:V16.0.0} presents the agreed positioning techniques, including the main signal and measurements considered at the UE and gNB side (i.e., downlink and uplink \ac{TDOA}, \ac{AOA}, \ac{RSRP}). More importantly, further standardization work is ongoing for enhanced localization services.  Therefore, in this paper, we present a class of approaches for the detection of anomalies due to intentional interference that aim at being general enough to be adapted to the current studied techniques and possible future techniques of the 5G \ac{NR}  as well as different types of location data (e.g., \ac{TDOA} and \ac{AOA}). %In what follows, we will generically refer to 
%location data as \emph{measurements} and assume that they are collected through a temporal window 
%of preassigned duration, which is used to track a slowly moving \ac{UE}. 
%In the case where intentional interference is not present, such measurements
%are statistically homogeneous, whereas if an attack takes place in the course of the window acquisition, the resulting measurements
%can be partitioned into two subsets representing the situation before and after the attack, respectively.

We first formulate the detection problem at hand in terms of a binary hypothesis test where data under the null hypothesis are homogeneous and 
distinguishing between either \ac{NLJ} or spoofing/meaconing attacks. 
At the design stage, we assume two statistical models for the available measurements that differ in the level 
of correlation among them. In order to solve the hypothesis test, we resort to both the plain \ac{GLRT} and to an {\em ad hoc}
modification that incorporates the \ac{LVM} \cite{murphy2012machine} (a point better explained in Section \ref{sec:nlj} and \ref{sec:spo}).
The performance analysis is conducted over both simulated and highlights the effectiveness
of the proposed approach. Finally, it is important to underline that such architectures can be thought as a component
of the location enablers of 5G that performs a preliminary analysis of data and provides a possible alert about the presence of
an attack. This alert is then jointly exploited by the mitigation and localization function to remedy possible information
counterfeit.

The remainder of the paper is organized as follows. Section \ref{sec:ProbForm} 
is devoted to sensor model description and to the formalization of the detection problems at hand, 
while the designs of the detection architectures are described in Section \ref{sec:nlj} and \ref{sec:spo} for \ac{NLJ} and spoofing attacks, respectively. 
In Section \ref{sec:performanceAnalysis}, some illustrative examples are provided to show
the effectiveness of the proposed strategies. 
Finally, Section \ref{sec:conclusions} contains concluding remarks and charts a course for future works.
	
\subsection{Notation} 
In the sequel, vectors and matrices are denoted by boldface lower-case and upper-case letters, respectively. 
Symbols $\det(\cdot)$, $\tr(\cdot)$, and $(\cdot)^T$ denote the determinant, trace, and transpose, respectively. 
%%%%If $A$ and $B$ are two generic sets, then $A\times B$ denotes the Cartesian product between $A$ and $B$.
As to the numerical sets, $\R$ is the set of real numbers and $\R^{N\times M}$ 
is the Euclidean space of $(N\times M)$-dimensional real matrices (or vectors if $M=1$).
%%%%$\C$ is the set 
%%%%of complex numbers, and $\C^{N\times M}$ is the Euclidean space of $(N\times M)$-dimensional complex matrices 
%%%%(or vectors if $M=1$).  The imaginary unit is indicated by $j$.
The Euclidean norm of a generic vector $\V{x}$ is denoted by $\|\V{x}\|$ whereas the modulus of a real number $x$ is denoted by $|x|$. 
%%%The symbol $\E[\cdot]$ denotes statistical expectation while $\V{0}$ is the null vector or matrix of proper size. 
%%%Given two events $A$ and $B$, the conditional probability of $A$ given $B$ is denoted by $P(A|B)$.
%%%For a given matrix $\bA$, $\lambda_{\max}\{\bA\}$ denotes the maximum eigenvalue of $\bA$. 
For any $N$-dimensional vector $\V{x}$, $\V{X}=\diag\{\V{x}\}$ is a $(N\times N)$-dimensional diagonal matrix whose principal diagonal
contains the elements of $\V{x}$.
Symbols $\V{I}$ and $\V{0}$ indicate the identity matrix and the null matrix or vector, respectively, whose size depends on the context.
The curled inequality symbol $\succeq$ (and its strict form $\succ$) is used to denote generalized matrix 
inequality: for any $N$-dimensional Hermitian matrix $\V{A}$, $\V{A}\succeq\V{0}$ means that $\V{A}$ is a positive 
semi-definite matrix ($\V{A}\succ\V{0}$ for positive definiteness).
Finally, we write $\V{x}\sim\cN_N(\V{m}, \V{M})$ if $\V{x}$ is a $N$-dimensional Gaussian vector 
with mean $\V{m}$ and covariance matrix $\V{M}\succ \V{0}$, whereas, given $\V{X}\in\R^{N\times K}$, 
$\V{X}\sim\cN_N(\V{m}, \V{M},\V{I})$ means that the columns of $\V{X}$ are \ac{IID} random vectors following the Gaussian
distribution with mean $\V{m}$ and covariance matrix $\V{M}$.

\section{Sensor Model and Problem Formulation}
\label{sec:ProbForm}
Consider a slowly moving \ac{UE} that is under tracking by the network infrastructure. 
Let us denote by $\V{Z}=[\V{z}_1, \V{z}_2, \ldots, \V{z}_K]\in\R^{N\times K}$ the entire data matrix
whose $k$th column, $k=1,\ldots,K$, contains a set of measurements acquired at the $k$th time istant.
For instance, such measurements can be \ac{DOA}, the \ac{TOA}, the \ac{OTDOA}, the \ac{RSRP}, or others provided by the network.

In what follows, we assume that the measurement errors are independent over the time and thdic
 
the Gaussian distribution with zero mean and covariance matrix depending upon the specific operating scenario (a point better
explained below). Thus, in a scenario unaffected by malicious platform actions, all the measurement are \ac{IID}, 
namely\footnote{Notice that we are also assuming that the unintentional interference is stationary
at least within the observation time interval.}
\begin{equation}
\label{eq:meas}
\V{z}_k\sim \cN_N(\V{m}, \M{\Sigma}), \quad k=1,\ldots,K,
\end{equation}
where $\V{m}\in\R^{N\times 1}$ contains the actual values of the considered location parameters and $\V{\Sigma}\in\R^{N\times N}$ is the
positive definite error covariance matrix, which either can exhibit a generic symmetric structure or can be diagonal. The former case allows to
accounts for a possible correlation among the measurements at the design stage, whereas the latter case corresponds to independent measurements.

Now, if at a certain time index, $K_0$ say, within the observation interval, a malicious platform performs an attack aimed at disrupting 
the receiver functionalities by transmitting noise-like signals, the quality of the estimates provided by the sensors 
would impair due to an increased uncertainty. As a consequence, data matrix can be partitioned into the following two submatrices
$\V{Z}_{1:K_{0}} \sim \cN_N(\V{m},\V{\Sigma}_1,\V{I})$ and 
$\V{Z}_{K_0+1:K} \sim \cN_N(\V{m},\V{\Sigma}_2,\V{I})$,
%\ee
where $\V{Z}_{1:K_{0}}=[\V{z}_1,\ldots,\V{z}_{K_0}]\in\R^{N\times K_0}$, 
$\V{Z}_{K_0+1:K}=[\V{z}_{K_0+1},\ldots,\V{z}_{K}]\in\R^{N\times (K-K_0)}$,
$\V{\Sigma}_2-\V{\Sigma}_1\succ \V{0}$, and $K_0\in\Omega_0\subseteq\Omega=\{1,\ldots,K\}$.
Under the above assumptions, the detection problem at hand can be formulated in terms of the following hypothesis test
\be 
\left\{
\begin{aligned}
& H_0:\V{Z}\sim \cN_N(\V{m}_0,\V{\Sigma}_0,\V{I}),
\\
& H_1:
\begin{cases}
\V{Z}_{1:K_0} \sim \cN_N(\V{m}_1,\V{\Sigma}_1,\V{I}),
\\
\V{Z}_{K_0+1:K} \sim \cN_N(\V{m}_1,\V{\Sigma}_2,\V{I}),
\end{cases}
\end{aligned}
\right.
\label{eq:hyptestNLJ}
\ee 
where $\V{m}$, $\V{\Sigma}_0\succ \V{0}$, $\V{\Sigma}_1\succ \V{0}$, $\V{\Sigma}_2\succ \V{0}$, and $K_0$ are unknown.

On the other hand, if the hostile platform is aimed at injecting false information into the network receivers, then, starting from the time
instant $K_0$, the mean vector no longer contains the true position of the \ac{UE} but its entries are related to the false 
position information created by the attacker. Therefore, in this case, the considered detection problem becomes
\be 
\left\{
\begin{aligned}
& H_0:\V{Z}\sim \cN_N(\V{m}_0,\V{\Sigma}_0,\V{I}),
\\
& H_1:
\begin{cases}
\V{Z}_{1:K_0} \sim \cN_N(\V{m}_1,\V{\Sigma}_1,\V{I}),
\\
\V{Z}_{K_0+1:K} \sim \cN_N(\V{m}_2,\V{\Sigma}_1,\V{I}),
\end{cases}
\end{aligned}
\right.
\label{eq:hyptestSpoof}
\ee
where $\V{m}_1\in\R^{N\times 1}$ is an unknown vector containing the location information before that 
spoofing takes place, $\V{m}_2\in\R^{N\times 1}$ is an unknown vector representing the modified position due to the attack,
$\V{\Sigma}_1\succ\V{0}$, and, again, $K_0$ is unknown.

In the next section, we devise different detection architectures for the above problems relying on the \ac{GLRT} and \ac{LVM}.

Before concluding this section, we provide some definitions which come in handy for the next developments.
More precisely, the \ac{PDF} of $\V{Z}$ under $H_0$ for both problems has the following 
expression $f_0(\V{Z};\V{m}_0,\!\V{\Sigma}_0) \!\!\!\!\! = \!\!\!\!\! \prod_{k=1}^K \! f_0(\V{z}_k;\V{m}_0,\!\V{\Sigma}_0)$,
while the \ac{PDF}s of $\V{Z}$ under $H_1$ for problems \eqref{eq:hyptestNLJ} and \eqref{eq:hyptestSpoof} are given by
$f_1(\V{Z};\V{m}_1,\V{\Sigma}_1,\V{\Sigma}_2,K_0)=\prod_{k=1}^{K_0} f_1(\V{z}_k;\V{m}_1,\V{\Sigma}_1)$ $\times\prod_{k=K_0+1}^{K} f_1(\V{z}_k;\V{m}_1,\V{\Sigma}_2)$ and $f_1(\V{Z};\V{m}_1,\V{m}_2,\V{\Sigma}_1,K_0)=
\prod_{k=1}^{K_0} \! f_1(\V{z}_k;\V{m}_1,\!\V{\Sigma}_1) \!\prod_{k=K_0+1}^{K}\! f_1(\V{z}_k;\V{m}_2,\!\V{\Sigma}_1)$,
respectively, where
\begin{multline}
f_l(\V{z}_k;\V{m}_m,\V{\Sigma}_n)
\\
=\frac{\exp\left\{\ds -\frac{1}{2}\tr[\V{\Sigma}_n^{-1}(\V{z}_k-\V{m}_m)(\V{z}_k-\V{m}_m)^T] \right\}}
{(2\pi)^{N/2} [\det(\V{\Sigma}_n)]^{1/2}}
\end{multline}
with $l=0,1$, $m=0,1,2$, and $n=0,1,2$.

%\section{}
%\label{sec:design}
%In this section, we exploit \ac{GLRT}-based design procedures to come up with adaptive decision schemes for problems \eqref{eq:hyptestNLJ}
%and \eqref{eq:hyptestSpoof} assuming different structures for the error covariance matrices. Moreover, for the most general covariance structure,
%we also exploit a classification procedure that jointly resorts to the \ac{EM} algorithm \cite{Dempster77} and the \ac{LVM} \cite{murphy2012machine}.

\section{Noise-like Jammer Detectors}
\label{sec:nlj}
Decision rules presented in this section are derived under the design assumptions of problem \eqref{eq:hyptestNLJ} assuming different structures for the error covariance matrices. Moreover, for the most general covariance structure,
%we also exploit a classification procedure that jointly resorts to the \ac{EM} algorithm \cite{Dempster77} and the \ac{LVM} \cite{murphy2012machine}.. Let us 
start with the plain \ac{GLRT} whose general structure is given by
\be
\frac{\dmax_{K_0\in\Omega_0}\dmax_{\V{m}_1}\dmax_{\V{\Sigma}_1}\dmax_{\V{\Sigma}_2}f_1(\V{Z};\V{m}_1,\V{\Sigma}_1,\V{\Sigma}_2,K_0)}
{\dmax_{\V{m}_0}\dmax_{\V{\Sigma}_0} f_0(\V{Z};\V{m}_0,\V{\Sigma}_0)}
\test \eta,
\label{eq:GLRT_iniNLJ}
\ee
where $\Omega_0\subseteq \{1,\ldots,K\}$ and $\eta$ is the detection 
threshold\footnote{Hereafter, we denote by $\eta$ the generic detection threshold.} 
to be set in order to ensure a preassigned Probability of False Alarm ($P_\text{fa}$), and consider two cases
\begin{itemize}
\item the available measurements are uncorrelated leading to diagonal covariance matrices;
\item there exists a correlation among the measurements, i.e., the covariance matrices are generally symmetric.
\end{itemize}

\subsection{GLRT for Uncorrelated Measurements}
\label{subsubsec:uncorrGLRT_jamm}
In this case, we assume that $\V{\Sigma}_0 =\diag\{ \sigma^2_{0,1},\ldots,\sigma^2_{0,N} \}$, 
$\V{\Sigma}_1 =\diag\{ \sigma^2_{1,1},\ldots,\sigma^2_{1,N} \}$, and 
$\V{\Sigma}_2 =\diag\{ \sigma^2_{2,1},\ldots,\sigma^2_{2,N} \}$ 
$=\diag\{ \sigma^2_{1,1}+\Delta\sigma^2_{1},\ldots,\sigma^2_{1,N}+\Delta\sigma^2_{N} \}$.
Now, in order to compute the left-hand side of \eqref{eq:GLRT_iniNLJ}, we begin with the optimization problems under $H_0$.
To this end, for computational convenience, we consider the log-likelihood function that, neglecting the constants irrelevant 
to the maximization, can be written as
\begin{multline}
\cL_0(\V{m}_0,\V{\Sigma}_0;\V{Z}) \approx 
-\frac{K}{2} \sum_{n=1}^{N} \log{\sigma_{0,n}^2} 
\\
-\frac{1}{2}\sum_{k=1}^{K} \sum_{n=1}^{N} \frac{(z_{k,n}- 
m_{0,n})^2}{\sigma_{0,n}^2},
\end{multline}
$m_{0,1},\ldots,m_{0,N}$ are the entries of $\V{m}_0$ and $z_{k,1},\ldots,z_{k,N}$ are the entries of $\V{z}_k$, $k=1,\ldots,K$.
Setting to zero the gradient of $\cL_0(\V{m}_0,\V{\Sigma}_0;\V{Z})$ with respect to $\V{m}_0$ and $\V{\Sigma}_0$, 
we come up with the following estimates for $n=1,\ldots,N$
\be
%\begin{cases}
\ds\widehat{m}_{0,n} = \frac{1}{K} \sum_{k=1}^{K}  z_{k,n}, \qquad \ds\widehat{\sigma}^2_{0,n} =\frac{1}{K} \sum_{k=1}^{K} (z_{k,n} - \widehat{m}_{0,n})^2,  
%\end{cases}
\label{eq:MLE_H0_NLJ_diag}
\ee
which are maximum points since, $\forall n=1,\ldots,N$, the following inequalities hold
\be
\begin{cases}
\forall {\sigma}^2_{0,n}\!\!<\!\!\ds\frac{1}{K} \sum_{k=1}^{K} (z_{k,n} - {m}_{0,n})^2:& \!\!\!\!\!\!\!
 \ds\frac{\partial}{\partial \sigma^2_{0,n}}[\cL_0(\V{m}_0,\V{\Sigma}_0;\V{Z})]>0,
\\
\forall {\sigma}^2_{0,n}\!\!>\!\!\ds\frac{1}{K} \sum_{k=1}^{K} (z_{k,n} - {m}_{0,n})^2:& 
\!\!\!\!\!\!\!
\ds\frac{\partial}{\partial \sigma^2_{0,n}}[\cL_0(\V{m}_0,\V{\Sigma}_0;\V{Z})]<0,
\\
\ds \forall {m}_{0,n} \!\!<\!\! \frac{1}{K} \sum_{k=1}^{K}  z_{k,n} : & \!\!\!\!\!\!\! 
\ds\frac{\partial}{\partial m_{0,n}}[\cL_0(\V{m}_0,\V{\Sigma}_0;\V{Z})]>0,
\\
\ds \forall {m}_{0,n} \!\!>\!\! \frac{1}{K} \sum_{k=1}^{K}  z_{k,n} :& \!\!\!\!\!\!\! 
\ds\frac{\partial}{\partial m_{0,n}}[\cL_0(\V{m}_0,\V{\Sigma}_0;\V{Z})]<0.
\end{cases}
\ee
As for the maximization under $H_1$, the corresponding log-likelihood function can be written (again neglecting the irrelevant constants)
as follows
\begin{multline}
\cL_1(\V{m}_1,\V{\Sigma}_1,\V{\Sigma}_2,K_0;\V{Z}) \approx -\frac{K_0}{2} \sum_{n=1}^{N} \log \sigma_{1,n}^2  
\\ 
-\frac{K_1}{2} \sum_{n=1}^{N}  \log (\sigma_{1,n}^2+ \Delta\sigma_{n}^2)- \frac{1}{2} \sum_{k=1}^{K_0} \sum_{n=1}^{N} 
\frac{(z_{k,n}-{m}_{1,n})^2}{ \sigma_{1,n}^2}
\\
-\frac{1}{2}\sum_{k=K_0+1}^{K}  \sum_{n=1}^{N} \frac{({z}_{k,n}-{m}_{1,n})^2}{ \sigma^2_{1,n}+\Delta\sigma^2_{n}},
\label{eq:LL_H1_NLJ_diag}
\end{multline}
where $K_1=K-K_0$. Before proceeding with the search of the stationary points of $\cL_1(\V{m}_1,\V{\Sigma}_1,\V{\Sigma}_2,K_0;\V{Z})$,
we compute the following limits that tell us where the stationary points can be sought, namely
\begin{align}
&\lim_{\|\V{m}\|\rightarrow+\infty}\cL_1(\V{m}_1,\V{\Sigma}_1,\V{\Sigma}_2,K_0;\V{Z})=-\infty, \nonumber
\\
&\lim_{\sigma_{1,n}^2\rightarrow+\infty}\cL_1(\V{m}_1,\V{\Sigma}_1,\V{\Sigma}_2,K_0;\V{Z})=-\infty, \ n=1,\ldots,N, \nonumber
\\
&\lim_{\sigma_{1,n}^2\rightarrow 0}\cL_1(\V{m}_1,\V{\Sigma}_1,\V{\Sigma}_2,K_0;\V{Z})=-\infty, \ n=1,\ldots,N, \nonumber
\\
&\lim_{\Delta\sigma_{n}^2\rightarrow+\infty}\cL_1(\V{m}_1,\V{\Sigma}_1,\V{\Sigma}_2,K_0;\V{Z})=-\infty, \ n=1,\ldots,N, \nonumber
\\
&\lim_{\Delta\sigma_{n}^2\rightarrow 0}\cL_1(\V{m}_1,\V{\Sigma}_1,\V{\Sigma}_2,K_0;\V{Z})=C\in\R, \ n=1,\ldots,N. \nonumber
\end{align}
Thus, we can search the stationary points in the interior of the log-likelihood domain by setting to zero the first derivatives with respect
to the unknown parameters. As first step, we consider the maximization with respect to $\sigma^2_{1,n}$ and $\Delta\sigma^2_{1,n}$, $n=1,\ldots,N$.
To this end, let us compute 
\be
\frac{\partial}{\partial \Delta\sigma^2_{1,n}}[\cL_1(\V{m}_1,\V{\Sigma}_1,\V{\Sigma}_2,K_0;\V{Z})]=0, \ n=1,\ldots,N,
\ee
which leads to the following equality for $n=1,\ldots,N$\footnote{From the sign of the derivative, it is possible to show that \eqref{eq:sigma2_MLE}
is a maximum point.}
\be
\widehat{\Delta\sigma^2_{n}}(\sigma_{1,n}^2,m_{1,n})=
\begin{cases}
\ds\frac{1}{K_1} \sum_{k=K_0+1}^{K} (z_{k,n}-m_{1,n})^2-\sigma_{1,n}^2
\\
\mbox{ if } \widehat{\Delta\sigma^2_{n}}>0,
\\
0, \mbox{ otherwise}.
\end{cases}
\label{eq:sigma2_MLE}
\ee
Now, let $\Gamma_N=\{1,\ldots,N\}$ and, given a value of $K_0$, define the following sets
\begin{align}
\Gamma(K_0) &=\{n\in\Gamma_N: \ \widehat{\Delta\sigma^2_{n}}(\sigma_{1,n}^2,m_{1,n})>0 \}\subseteq \Gamma_N,
\\
\bar{\Gamma}(K_0) &=\{n\in\Gamma_N: \ \widehat{\Delta\sigma^2_{n}}(\sigma_{1,n}^2,m_{1,n})=0 \}\subseteq \Gamma_N.
\end{align}
Thus, replacing the above stationary points in \eqref{eq:LL_H1_NLJ_diag} and exploiting the last two definitions, 
we obtain (neglecting the constants)
a
\vspace{-0.4cm}
\begin{multline}
\cL_1(\V{m}_1,\V{\Sigma}_1,\widehat{\V{\Sigma}}_2,K_0;\V{Z}) \approx -\frac{K_0}{2} \sum_{n=1}^{N} \log \sigma_{1,n}^2  
\\ 
-\frac{K_1}{2} \sum_{n\in\Gamma(K_0)}  \log \left[\frac{1}{K_1} \sum_{k=K_0+1}^{K} (z_{k,n}-m_{1,n})^2\right]
\\
-\frac{K_1}{2} \sum_{n\in\bar{\Gamma}(K_0)}  \log \sigma_{1,n}^2
- \frac{1}{2}  \sum_{n=1}^{N} \sum_{k=1}^{K_0}
\frac{(z_{k,n}-{m}_{1,n})^2}{ \sigma_{1,n}^2}
\\
- \frac{1}{2}  \sum_{n\in\bar{\Gamma}(K_0)} \sum_{k=K_0+1}^{K}
\frac{(z_{k,n}-{m}_{1,n})^2}{ \sigma_{1,n}^2}.
\label{eq:LL_H1_NLJ_diag_01}
\end{multline}
Let us proceed by setting to zero the first derivative of $\cL_1(\V{m}_1,\V{\Sigma}_1,\widehat{\V{\Sigma}}_2,K_0;\V{Z})$ 
with respect to $\sigma_{1,n}$, $n\in\Gamma(K_0)\cup \bar{\Gamma}(K_0)$, to obtain
\begin{align}
\widehat{\sigma}_{1,n}^2(m_{1,n})&= \frac{1}{K_0} \sum_{k=1}^{K_0} (z_{k,n}-m_{1,n})^2, \ n\in\Gamma(K_0), \nonumber
\\
\label{eqn:sigma1_estimate}
\widehat{\sigma}_{1,n}^2(m_{1,n})&= \frac{1}{K} \sum_{k=1}^{K} (z_{k,n}-m_{1,n})^2, \ n\in\bar{\Gamma}(K_0).
\end{align}
It follows that the compressed log-likelihood function can be written as
\begin{multline}
\cL_1(\V{m}_1,\widehat{\V{\Sigma}}_1,\widehat{\V{\Sigma}}_2,K_0;\V{Z}) 
\\
\approx -\frac{K_0}{2} \sum_{n\in\Gamma(K_0)} 
\log \left[ \frac{1}{K_0} \sum_{k=1}^{K_0} (z_{k,n}-m_{1,n})^2 \right]
\\ 
-\frac{K_1}{2} \sum_{n\in\Gamma(K_0)}  \log \left[\frac{1}{K_1} \sum_{k=K_0+1}^{K} (z_{k,n}-m_{1,n})^2\right]
\\ 
-\frac{K}{2} \sum_{n\in\bar{\Gamma}(K_0)}  \log \left[\frac{1}{K} \sum_{k=1}^{K} (z_{k,n}-m_{1,n})^2\right].
\label{eq:LL_H1_NLJ_diag_02}
\end{multline}
It still remains to maximize the log-likelihood function with respect to $\V{m}_1$ and $K_0$. To this end, we observe that
optimization over $K_0$ can be carried out through a $1$-dimensional grid search, whereas an estimate of $\V{m}_1$
can be obtained by solving the following system of equations
\be
\frac{\partial}{\partial m_{1,n}}[\cL_1(\V{m}_1,\widehat{\V{\Sigma}}_1,\widehat{\V{\Sigma}}_2,K_0;\V{Z})]=0, \ n\in\Gamma(K_0)\cup \bar{\Gamma}(K_0).
\ee
When $n\in\bar{\Gamma}(K_0)$, it is not difficult to show that the corresponding equation leads to $\widehat{m}_{0,n}$ given by
\eqref{eq:MLE_H0_NLJ_diag}, whereas in the case where $n\in{\Gamma}(K_0)$, we have to solve
\begin{align}
&\frac{\ds\sum_{k=1}^{K_0} z_{k,n} - K_0 m_{1,n}}{\ds\frac{1}{K_0} \sum_{k=1}^{K_0} (z_{k,n}-m_{1,n})^2} 
+  \frac{\ds\sum_{k=K_0+1}^{K} z_{k,n} - K_1 m_{1,n}}{\ds\frac{1}{K_1} \sum_{k=K_0+1}^{K}(z_{k,n}-m_{1,n})^2}=0 \nonumber
\\
&\Rightarrow 
\frac{1}{K_1} \sum_{k=K_0+1}^{K}(z_{k,n}-m_{1,n})^2 \left[ \sum_{k=1}^{K_0} z_{k,n} - K_0 m_{1,n} \right] \nonumber
\\
& + \frac{1}{K_0} \sum_{k=1}^{K_0} (z_{k,n}-m_{1,n})^2\left[ \sum_{k=K_0+1}^{K} z_{k,n} - K_1 m_{1,n} \right]=0.
\label{eq:m1_est_eq}
\end{align}
Now, let us define the following quantities
$A_{K_0,n} = \sum_{k=1}^{K_0} z_{k,n}$, $\quad B_{K_0,n} = \sum_{k=1}^{K_0} z_{k,n}^2$,
$A_{K_1,n} = \sum_{k=K_0+1}^{K} z_{k,n}$, and $\quad B_{K_1,n} = \sum_{k=K_0+1}^{K} z_{k,n}^2$,
then, \eqref{eq:m1_est_eq} can be recast as
\begin{align}
&\sum_{i,j\in \{0,1\} \atop i\neq j}\frac{[B_{K_i,n}+K_i m_{1,n}^2-2m_{1,n}A_{K_i,n}][A_{K_j,n}-K_j m_{1,n}]}{K_i} \nonumber
\\
&\Rightarrow C_3 m_{1,n}^3+C_2 m_{1,n}^2+C_1 m_{1,n}+C_0=0,
\end{align}
where $C_0 = \frac{B_{K_1,n} A_{K_0,n} }{K_1}  +  \frac{B_{K_0,n} A_{K_1,n} }{K_0}$, 
$C_1 = - \frac{K_0}{K_1} B_{K_1,n} - \frac{K_1}{K_0} B_{K_0,n} - \left(\frac{2}{K_1} + \frac{2}{K_0}  \right)A_{K_1,n} A_{K_0,n}$,
$C_2 = A_{K_0,n}+A_{K_1,n} +2 \frac{K_0}{K_1}A_{K_1,n}+2 \frac{K_1}{K_0} A_{K_0,n}$, and
$C_3 = -(K_0+K_1)$.
The solutions of the above equations can be explicitly obtained resorting to Cardano's method \cite{childs2008concrete} and, then, 
we choose that one, $\widehat{\V{m}}_{1}$ say, leading to the maximum of 
$\cL_1(\V{m}_1,\widehat{\V{\Sigma}}_1,\widehat{\V{\Sigma}}_2,K_0;\V{Z})$, for all admissible 
values of\footnote{In the case of uncorrelated measurements, $K_0$ is not subject to any constraint and, hence, it takes on value 
in $\Omega$} $K_0\in\Omega$.

Finally, notice that in order to estimate $\Gamma(K_0)$ (and, hence, $\bar{\Gamma}(K_0)$), 
we resort to the inequality in \eqref{eq:sigma2_MLE} after having replaced the unknown parameters with the respective estimates, namely,
we verify that
\be
\frac{1}{K_1} \sum_{k=K_0+1}^{K} (z_{k,n}-\widehat{m}_{1,n})^2
-\frac{1}{K_0} \sum_{k=1}^{K_0} (z_{k,n}-\widehat{m}_{1,n})^2 > 0.
\ee
In what follows, with a little abuse of notation, we maintain symbols $\Gamma(K_0)$ and $\bar{\Gamma}(K_0)$ for their respective estimates.

Gathering the above results, the logarithm of the \ac{GLRT} can be written as
\begin{multline}
\dmax_{K_0}\Bigg\{
-\frac{K_0}{2} \sum_{n\in\Gamma(K_0)} 
\log \left[ \frac{1}{K_0} \sum_{k=1}^{K_0} (z_{k,n}-\widehat{m}_{1,n})^2 \right] 
\\
-\frac{K_1}{2} \sum_{n\in\Gamma(K_0)}  \log \left[\frac{1}{K_1} \sum_{k=K_0+1}^{K} (z_{k,n}-\widehat{m}_{1,n})^2\right]
\\
+\frac{K}{2} \sum_{n\in\Gamma(K_0)} \log \left[ \frac{1}{K} \sum_{k=1}^{K} (z_{k,n}-\widehat{m}_{0,n})^2 \right]\Bigg\}\test \eta.
\end{multline}
Note that if $\Gamma(K_0)=\emptyset$, then the decision statistic is equal to zero. 
In the following, we refer to these
decision rule as \ac{NLJ} Detector for Uncorrelated Measurements (NLJ-D-UM).

\subsection{GLRT for Correlated Measurements}
In the presence of correlated measurements, the covariance matrices $\V{\Sigma}_i$, $i=0,1,2$, in \eqref{eq:hyptestNLJ} are no longer 
diagonal but positive definite symmetric. 
Moreover, we assume that $K_1\geq N$ and $K_0\geq N$. These constraints ensure that the sample covariance matrices based upon
$\V{Z}_{1:K_0}$ and $\V{Z}_{K_01:K}$ are nonsingular with probability $1$ \cite{muirhead2009aspects}. As a consequence,
in this case $\Omega_0=\{N,N+1,\ldots,K-N\}$. It is important to highlight that from an operating point of view and for a sufficiently 
wide sliding window moving along the temporal dimension, the above requirements may be fulfilled.

Under these design assumptions, the denominator of the left-hand side of \eqref{eq:GLRT_iniNLJ}
can be easily simplified replacing $\V{m}_0$ and $\V{\Sigma}_0$ with their respective \ac{MLE}, given by 
$\bar{\V{m}}_0 = \frac{1}{K} \sum_{k=1}^{K} \V{z}_k$ and 
$\bar{\V{\Sigma}}_0 = \frac{1}{K} \sum_{k=1}^{K} (\V{z}_k-\bar{\V{m}}_0)(\V{z}_k-\bar{\V{m}}_0)^T$ (see, e.g. \cite{muirhead2009aspects}).
On the other hand, under $H_1$, the log-likelihood function can be written as (up to constants)
\begin{multline}
\cL_1(\V{m}_1,\V{\Sigma}_1,\V{\Sigma}_2,K_0;\V{Z})\approx
- \frac{K_0}{2} \log \det(\V{\Sigma}_1)
\\
- \frac{K_1}{2} \log \det (\V{\Sigma}_1+\V{R}) 
- \frac{1}{2} \sum_{k=1}^{K_0} (\V{z}_k-\V{m}_1)^T \V{\Sigma}_1^{-1} (\V{z}_k-\V{m}_1)
\\
- \frac{1}{2} \sum_{k=K_0+1}^{K} (\V{z}_k-\V{m}_1)^T (\V{\Sigma}_1+\V{R})^{-1}(\V{z}_k-\V{m}_1),
\end{multline}
where $\V{R}=\V{\Sigma}_2-\V{\Sigma}_1\succ\V{0}$. Now, notice that, for a preassigned $\V{\Sigma}_1$, the transformation
$\V{\Sigma}_2=\V{\Sigma}_1+\V{R}$ is one-to-one. For this reason, the maximum of $\cL_1(\V{m}_1,\V{\Sigma}_1,\V{\Sigma}_2;\V{Z})$
with respect to $\V{R}$ can be attained at \cite{muirhead2009aspects}
\begin{align}
\bar{\V{\Sigma}}_2=\frac{1}{K_1}\sum_{k=K_0+1}^{K}(\V{z}_k-\V{m}_1)(\V{z}_k-\V{m}_1)^T.
\end{align}
It follows that
\be
\bar{\V{R}}(\V{\Sigma}_1,\V{m}_1)=
\begin{cases}
\ds\frac{1}{K_1}\sum_{k=K_0+1}^{K}(\V{z}_k-\V{m}_1)(\V{z}_k-\V{m}_1)^T-\V{\Sigma}_1 
\\ \mbox{ if } \bar{\V{R}}\succ \V{0},
\\
\V{0}, \quad \mbox{otherwise}. 
\end{cases}
\ee
In the case $\V{R}=\V{0}$, the remaining unknown parameters are estimated as under $H_0$ and final statistic becomes 
a constant equal to zero. In the opposite case, we proceed by optimizing
\begin{multline}
\cL_1(\V{m}_1,\V{\Sigma}_1,\bar{\V{\Sigma}}_2,K_0;\V{Z})\approx
- \frac{K_0}{2} \log \det(\V{\Sigma}_1)
\\
- \frac{K_1}{2} \log \det (\widehat{\V{\Sigma}}_2) 
- \frac{1}{2} \sum_{k=1}^{K_0} (\V{z}_k-\V{m}_1)^T \V{\Sigma}_1^{-1} (\V{z}_k-\V{m}_1),
\end{multline}
with respect to $\V{\Sigma}_1$ to obtain 
%\be
$\bar{\V{\Sigma}}_1(\V{m}_1) = \frac{1}{K_0}\sum_{k=1}^{K_0} (\V{z}_k-\V{m}_1) (\V{z}_k-\V{m}_1)^T$ (see \cite{muirhead2009aspects}).
%\ee$
After replacing the above estimate into the log-likelihood function under $H_1$, the latter becomes
\begin{align}
&\cL_1(\V{m}_1,\bar{\V{\Sigma}}_1,\bar{\V{\Sigma}}_2,K_0;\V{Z}) \nonumber
\\
&\approx - \frac{K_0}{2} \log \det\left[\frac{1}{K_0}\sum_{k=1}^{K_0} (\V{z}_k-\V{m}_1) (\V{z}_k-\V{m}_1)^T\right] \nonumber
\\
&- \frac{K_1}{2} \log \det \left[  \frac{1}{K_1}\sum_{k=K_0+1}^{K} \!\!\!\!\! (\V{z}_k-\V{m}_1)(\V{z}_k-\V{m}_1)^T \right] \nonumber
\\
&=- \frac{K_0}{2} \log \det\left[\sum_{k=1}^{K_0} (\V{z}_k-\V{m}_1) (\V{z}_k-\V{m}_1)^T\right]  \nonumber
\\
&- \frac{K_1}{2} \log \det \left[ \sum_{k=K_0+1}^{K} \!\!\!\!\! (\V{z}_k-\V{m}_1)(\V{z}_k-\V{m}_1)^T \right] \nonumber
\\
&+\frac{K_0 N}{2}\log K_0 +\frac{K_1 N}{2}\log K_1.
\end{align}
Let us focus on the terms depending on $\V{m}_1$ and define
\begin{multline}
g(\V{m}_1)=- \frac{K_0}{2} \log \det\left[\sum_{k=1}^{K_0} (\V{z}_k-\V{m}_1) (\V{z}_k-\V{m}_1)^T\right]
\\
- \frac{K_1}{2} \log \det \left[ \sum_{k=K_0+1}^{K} \!\!\!\!\! (\V{z}_k-\V{m}_1)(\V{z}_k-\V{m}_1)^T \right].
\end{multline}
The above function can be further recast as
\begin{align}
&g(\V{m}_1)= -\frac{K_0}{2}\log\det\left(   
\V{S}_0-\V{s}_0\V{m}_1^T-\V{m}_1\V{s}_0^T+K_0\V{m}_1\V{m}_1^T \right) \nonumber
\\
&-\frac{K_1}{2}\log\det\left(   
\V{S}_1-\V{s}_1\V{m}_1^T-\V{m}_1\V{s}_1^T+K_1\V{m}_1\V{m}_1^T \right) 
\\
&=-\frac{K_0}{2}\log\det\left(   
\V{M}_0 + \V{u}_0\V{u}_0^T \right) -\frac{K_1}{2}\log\det\left(\V{M}_1 + \V{u}_1\V{u}_1^T \right)
\label{eq:gm1_max}
\end{align}
where $\V{S}_0=\sum_{k=1}^{K_0} \V{z}_k\V{z}_k^T$, $\quad \V{S}_1=\sum_{k=K_0+1}^K \V{z}_k\V{z}_k^T$, $\V{s}_0=\sum_{k=1}^{K_0} \V{z}_k, \quad \V{s}_1=\sum_{k=K_0+1}^K \V{z}_k$,
\begin{align}
%& \nonumber
%\\
&\V{M}_0=\V{S}_0-\frac{1}{K_0}\V{s}_0\V{s}_0^T, \quad \V{M}_1=\V{S}_1-\frac{1}{K_1}\V{s}_1\V{s}_1^T, \nonumber
\\
&\V{u}_0=\frac{1}{\sqrt{K_0}}\V{s}_0-\sqrt{K_0}\V{m}_1, \quad \V{u}_1=\frac{1}{\sqrt{K_1}}\V{s}_1-\sqrt{K_1}\V{m}_1. \nonumber
\end{align}
Since $N\leq \min\{K_0, K_1\}$, both $\V{M}_0$ and $\V{M}_1$ are positive definite with probability $1$ and, hence, $g(\V{m}_1)$ becomes
\begin{align}
&g(\V{m}_1) = -\frac{K_0}{2}\log\det(\V{M}_0)-\frac{K_1}{2}\log\det(\V{M}_1) \nonumber
\\
&-\frac{K_0}{2}\log\det\left(   
\V{I} + \V{M}_0^{-1/2}\V{u}_0\V{u}_0^T\V{M}_0^{-1/2} \right) \nonumber
\\
&-\frac{K_1}{2}\log\det\left(\V{I} + \V{M}_1^{-1/2}\V{u}_1\V{u}_1^T\V{M}_1^{-1/2} \right)\nonumber
\\
&= -\frac{K_0}{2}\log\det(\V{M}_0)-\frac{K_1}{2}\log\det(\V{M}_1) \nonumber
\\ \label{eqn:g1m}
&-\frac{K_0}{2}\log\left(   
1 + \V{u}_0^T\V{M}_0^{-1}\V{u}_0 \right) -\frac{K_1}{2}\log\left(1 + \V{u}_1^T\V{M}_1^{-1}\V{u}_1 \right),  
\end{align}
where the last equality comes for the fact that $\forall \V{A}\in\R^{N\times M}$, $\V{B}\in\R^{M\times N}$, the equation
%\be
$\det(\V{I}+\V{A}\V{B})=\det(\V{I}+\V{B}\V{A})$ holds.
%\ee
The right-hand side of \eqref{eqn:g1m} clearly unveils the radially unbounded nature of $g(\V{m}_1)$, thus we can search the maximum points
in the interior of its domain by setting to zero its first derivative, namely $\frac{\partial}{\partial \V{m}_1} [g(\V{m}_1)]=\V{0}$. It follows that
\begin{align}
%& \nonumber
%\\
&\Rightarrow 2\V{M}_0^{-1} (\V{s}_0 - K_0 \V{m}_1)+2\V{M}_1^{-1} (\V{s}_1 - K_1  \V{m}_1)=0 \nonumber
\\
&\Rightarrow \bar{\V{m}}_1=\left( K_0\V{M}_0^{-1}+K_1\V{M}_1^{-1} \right)^{-1}
\left( \V{M}_0^{-1}\V{s}_0+ \V{M}_1^{-1}\V{s}_1 \right).\nonumber
\end{align}
It is also straightforward that the Hessian is negative definite, in fact, recalling that $\V{M}_i^{-1}\succ\V{0}$, $i=0,1$,
it turns out that
\be
\frac{\partial^2}{\partial \V{m}_1\partial \V{m}_1^T}[g(\V{m}_1)]=-2K_0 \V{M}_0^{-1} -2K_1 \V{M}_1^{-1} \prec \V{0}.
\ee
Gathering all the above results, the final expression for the logarithm of the GLRT is
\begin{multline}
\dmax_{K_0}\Bigg\{- \frac{K_0}{2} \log \det\left[\sum_{k=1}^{K_0} (\V{z}_k-\bar{\V{m}}_1) (\V{z}_k-\bar{\V{m}}_1)^T\right]
\\
- \frac{K_1}{2} \log \det \left[ \sum_{k=K_0+1}^{K} \!\!\!\!\! (\V{z}_k-\bar{\V{m}}_1)(\V{z}_k-\bar{\V{m}}_1)^T \right]\Bigg\}
\\
+\frac{K}{2} \log \det \left[ \sum_{k=1}^{K}  (\V{z}_k-\bar{\V{m}}_0)(\V{z}_k-\bar{\V{m}}_0)^T \right]\test \eta,
\label{eqn:GLRT_jammer_corr_final}
\end{multline}
when\footnote{The constants have been incorporated into the threshold.}
\begin{multline}
\frac{1}{K_1}\sum_{k=K_0+1}^{K}(\V{z}_k-\bar{\V{m}}_1)(\V{z}_k-\bar{\V{m}}_1)^T 
\\
- \frac{1}{K_0}\sum_{k=1}^{K_0} (\V{z}_k-\bar{\V{m}}_1) (\V{z}_k-\bar{\V{m}}_1)^T\succ \V{0}.
\end{multline}
In the case where the last condition does not hold true, the left-hand side of \eqref{eqn:GLRT_jammer_corr_final} is set to zero.

In what follows, we refer to these
decision rule as \ac{NLJ} Detector for Correlated Measurements (NLJ-D-CM).

\subsubsection{\ac{LVM} for Correlated Measurements}
\label{subsec:LVM_corr_NLJ}
Let us assume that the error covariance matrix exhibits a generic symmetric structure and, under $H_1$, introduce $K$ \ac{IID} 
discrete random variables, $\omega_k$ say, whose alphabet and unknown \ac{PMF} are $\cA=\{1,2\}$ and 
%\be
$P(\omega_k=a)=\pi_a, \quad a\in \cA, \quad k\in\Omega$, 
%\ee
respectively. Moreover, the $\omega_k$s are such that when $\omega_k=a$, then $\V{z}_k\sim\cN_N(\V{m}_1,\V{\Sigma}_a)$.
Therefore, the \ac{PDF} of $\V{z}_k$ can be written exploiting the Total Probability Theorem as
\be
f_1(\V{z}_k;\V{\pi},\V{m}_1,\V{\Sigma}_1,\V{\Sigma}_2)=\sum_{a\in\cA} \pi_a f_1(\V{z}_k;\V{m}_1,\V{\Sigma}_a),
\label{eq:pdf_LVM_NLJ}
\ee
where $\V{\pi}=[\pi_1 \ \pi_2]^T$. 
It is important to observe that the \ac{PDF} of $\V{Z}$ no longer depends on $K_0$. Now, applying the maximum 
likelihood approach to obtain the closed-form expressions for estimates of the unknown parameters leads to 
mathematically difficult optimization problems (at least to the best of authors knowledge).
For this reason, we resort to the \ac{EM} algorithm  that provides closed-form updates for the parameter estimates at each step and 
reaches at least a local stationary point. To this end, let us begin by writing the joint log-likelihood function of $\V{Z}$ (assuming
that \eqref{eq:pdf_LVM_NLJ} holds)
\begin{align}
&\log f_1(\V{Z};\V{\pi},\V{m}_1,\V{\Sigma}_1,\V{\Sigma}_2) \nonumber
\\
&=\sum_{k=1}^K \log\left[ \sum_{a\in\cA} \pi_a f_1(\V{z}_k;\V{m}_1,\V{\Sigma}_a) \right] \nonumber
\\
&=\sum_{k=1}^K \log\left[ \sum_{a\in\cA} q_k(a)\frac{ \pi_a f_1(\V{z}_k;\V{m}_1,\V{\Sigma}_a)}{q_k(a)} \right] \nonumber
\\
&\geq \sum_{k=1}^K \sum_{a\in\cA} \log\left[  \frac{ \pi_a f_1(\V{z}_k;\V{m}_1,\V{\Sigma}_a)}{q_k(a)} \right] q_k(a),
\end{align}
where $q_k(a)$ is an arbitrary \ac{PMF} over $\omega_k$ and the last inequality is due to Jensen inequality \cite{cover2012elements};
the equality holds when
\be
q_k(a)=\frac{ \pi_a f_1(\V{z}_k;\V{m}_1,\V{\Sigma}_a)}{\ds\sum_{m\in\cA} \pi_m f_1(\V{z}_k;\V{m}_1,\V{\Sigma}_m)}.
\ee
Assuming that at the $(h-1)$th step the estimates $\tilde{\V{\pi}}^{(h-1)}$, $\tilde{\V{m}}_1^{(h-1)}$, $\tilde{\V{\Sigma}}_1^{(h-1)}$, 
and $\tilde{\V{\Sigma}}_2^{(h-1)}$ are available, the updates related to the E-step are given by
\be
\tilde{q}^{(h-1)}_k(a)=\frac{ \tilde{{\pi}}_a^{(h-1)} f_1(\V{z}_k;\tilde{\V{m}}_1^{(h-1)},\tilde{\V{\Sigma}}^{(h-1)}_a)}
{\ds\sum_{m\in\cA} \tilde{{\pi}}_m^{(h-1)} f_1(\V{z}_k;\tilde{\V{m}}_1^{(h-1)},\tilde{\V{\Sigma}}^{(h-1)}_m)}.
\ee
Now, we proceed with the M-step which requires to solve
\begin{align}
&\dmax_{\V{\pi}}\dmax_{\V{m}_1}\dmax_{\V{\Sigma}_1}\dmax_{\V{\Sigma}_2}\Bigg\{
\sum_{k=1}^K \sum_{a\in\cA} \tilde{q}^{(h-1)}_k(a)\log(  { f_1(\V{z}_k;\V{m}_1,\V{\Sigma}_a)}) \nonumber
\\
&+\sum_{k=1}^K \sum_{a\in\cA} \tilde{q}^{(h-1)}_k(a)\log(\pi_a) \Bigg\}.
\end{align}
The maximization over $\V{\pi}$ can be accomplished by using the method of the Lagrange multipliers since $\pi_1+\pi_2=1$. 
Thus, the update for the prior estimate is given by
\be
\tilde{\pi}_a^{(h)}=\frac{1}{K}\sum_{k=1}^K \tilde{q}^{(h-1)}_k(a).
\label{eq:priorEstSpoof}
\ee
The last problems to be solved are
\be
\dmax_{\V{m}_1}\dmax_{\V{\Sigma}_a \atop a\in\cA}
\sum_{k=1}^K \sum_{a\in\cA} \tilde{q}^{(h-1)}_k(a)\log(  { f_1(\V{z}_k;\V{m}_1,\V{\Sigma}_a)}).
\label{eq:LVM_MstepNLJ}
\ee
In order to find the maximum with respect to $\V{\Sigma}_a$,
we replace the condition ${\V{\Sigma}}_2\succ {\V{\Sigma}}_1$ (this modification leads to more tractable solutions from
a mathematical point of view) and recast the objective function as
\begin{multline}
\sum_{a\in\cA} \frac{q^{(h-1)}(a)}{2}
\Bigg\{
\log\det(\V{\Sigma}_a^{-1})
\\
-\tr\left[   
\V{\Sigma}_a^{-1}\frac{1}{q^{(h-1)}(a)}\V{S}_a(\V{m}_1)
\right]
\Bigg\}
\end{multline}
where $q^{(h-1)}(a)=\sum_{k=1}^K \tilde{q}_k^{(h-1)}(a)$ and $\V{S}_a(\V{m}_1)=\sum_{k=1}^K \tilde{q}_k^{(h-1)}(a) 
(\V{z}_k-\V{m}_1)(\V{z}_k-\V{m}_1)^T$. Then, it is not difficult to show that
the maximization over $\V{\Sigma}_a$ leads to
\be
\tilde{\V{\Sigma}}_a^{(h)}=\frac{1}{q^{(h-1)}(a)}\V{S}_a(\V{m}_1), \quad a\in\cA.%, \mbox{ if } \tilde{\V{\Sigma}}_2^{(h)}\succeq \tilde{\V{\Sigma}}_1^{(h)}.
\ee
%In the case where $\tilde{\V{\Sigma}}_2^{(h)}\succeq \tilde{\V{\Sigma}}_1^{(h)}$ is not true, we set $\Lambda_1(\V{Z})=1$.
Finally, %assuming $\tilde{\V{\Sigma}}_2^{(h)}\succeq \tilde{\V{\Sigma}}_1^{(h)}$ and 
replacing the above estimates 
into \eqref{eq:LVM_MstepNLJ}, we obtain
\begin{align}
\dmax_{\V{m}_1}\left\{
-\sum_{a\in\cA} \frac{q^{(h-1)}(a)}{2}
\left[
\log\det\left(\frac{1}{q^{(h-1)}(a)}\V{S}_a(\V{m}_1)\right)+N
\right]\right\},
\end{align}
which can be solved by applying a suitable modification of the line of reasoning pursued for \eqref{eq:gm1_max}.
Thus, the estimate of $\V{m}_1$ has the following expression
\begin{multline}
\tilde{\V{m}}_1^{(h)}=\left(q^{(h-1)}(1) \V{M}_{0,1}^{-1}+q^{(h-1)}(2) \V{M}_{1,2}^{-1}\right)^{-1}
\\
\times \left(\V{M}_{0,1}^{-1}\V{s}_{0,1}+\V{M}_{1,2}^{-1}\V{s}_{1,2}\right),
\end{multline}
where
\begin{align}
&\V{M}_{0,1}=\sum_{k=1}^{K} \tilde{q}_k^{(h-1)}(1) \V{z}_k\V{z}_k^T-\frac{1}{q^{(h-1)}(1)}\V{s}_{0,1}\V{s}_{0,1}^T, \nonumber
\\
&\V{M}_{1,2}=\sum_{k=1}^{K} \tilde{q}_k^{(h-1)}(2) \V{z}_k\V{z}_k^T-\frac{1}{q^{(h-1)}(2)}\V{s}_{1,2}\V{s}_{1,2}^T, \nonumber
\\
&\V{s}_{0,1}=\sum_{k=1}^{K} \tilde{q}_k^{(h-1)}(1) \V{z}_k, \quad \V{s}_{1,2}=\sum_{k=1}^K \tilde{q}_k^{(h-1)}(2) \V{z}_k. \nonumber
\end{align}
The iterations of the EM algorithm terminate when a stopping criterion is satisfied. The latter 
can be related to the maximum number of iterations (dictated by computational power)
and/or the variation of the likelihood function at each iteration.

As for the maximization under $H_0$, since the latter is unaltered with respect to the previous case, we can clearly 
use the \ac{MLE} $\bar{\V{m}}_0$ and 
$\bar{\V{\Sigma}}_0$ and the final decision rule can be written as
\be
\label{eq:NLJLVM}
\frac{\ds\prod_{k=1}^K\sum_{a\in\cA} \tilde{\pi}_a^{(h)} f_1(\V{z}_k;\tilde{\V{m}}_1^{(h)},\tilde{\V{\Sigma}}_a^{(h)})}
{\ds f_0(\V{Z};\bar{\V{m}}_0,\bar{\V{\Sigma}}_0)}\test\eta.
\ee
The above decision rule will be referred to as LVM-based NLJ Detector (LVM-NLJ-D).

\section{Spoofing Detection Architectures}\label{sec:spo}
Recall that in the presence of a spoofing attack, the decision problem to be addressed is \eqref{eq:hyptestSpoof} and, hence,
the \ac{GLRT} has the following form
\be
\frac{\dmax_{K_0\in\Omega_0}\dmax_{\V{m}_1}\dmax_{\V{m}_2}\dmax_{\V{\Sigma}_1}
f_1(\V{Z};\V{m}_1,\V{m}_2,\V{\Sigma}_1,K_0)}
{\dmax_{\V{m}_0}\dmax_{\V{\Sigma}_0} f_0(\V{Z};\V{m}_0,\V{\Sigma}_0)}
\test \eta.
\label{eq:GLRT_ini}
\ee
In the next subsections, we derive the \ac{GLRT} for both correlated as well as uncorrelated measurements. Finally, we suitably modify 
it by incorporating the \ac{LVM} for the case of correlated measurements.

\subsection{GLRT for Uncorrelated Measurements}
Let us assume that the covariance matrix is diagonal, namely $\V{\Sigma}_0=\diag\{ \sigma_{0,1}^2,\sigma_{0,2}^2,\ldots,\sigma_{0,N}^2\}$, 
then, under $H_0$, the likelihood can be maximized resorting to \eqref{eq:MLE_H0_NLJ_diag}. 

On the other, under $H_1$, 
the resulting log-likelihood function has the following expression (neglecting the irrelevant constants)
\begin{multline}
\cL_1(\V{m}_1,\V{m}_2,\V{\Sigma}_1,K_0;\V{Z}) \approx -\frac{K}{2} \sum_{n=1}^{N} \log \sigma_{1,n}^2  
\\ 
- \frac{1}{2} \sum_{k=1}^{K_0} \sum_{n=1}^{N} 
\frac{(z_{k,n}-{m}_{1,n})^2}{ \sigma_{1,n}^2}
\\
-\frac{1}{2}\sum_{k=K_0+1}^{K}  \sum_{n=1}^{N} \frac{({z}_{k,n}-{m}_{2,n})^2}{ \sigma^2_{1,n}},
\label{eq:LL_H1_Spoof_diag}
\end{multline}
where $m_{i,n}$, $i=1,2$ and $n=1,\ldots,N$, is the $n$th entry of $\V{m}_i$.
Now, since $\cL_1(\V{m}_1,\V{m}_2,\V{\Sigma}_1,K_0;\V{Z})$ is radially unbounded with respect to $\|\V{m}_i\|$, $i=1,2$, we
can compute the stationary points by setting to zero its derivatives with respect to $m_{i,n}$, $n=1,\ldots,N$, 
obtaining the following estimates
\be
\widehat{m}_{1,n} =\ds\frac{1}{K_0}\sum_{k=1}^{K_0} z_{k,n}, \quad \widehat{m}_{2,n} =\ds\frac{1}{K_1}\sum_{k=K_0+1}^{K} z_{k,n}.
\ee
As for the estimates of $\V{\Sigma}_1$, let us define $\V{\sigma}_1=[\sigma_{1,1},\ldots,\sigma_{1,N}]^T$ and observe that
\begin{align}
\lim_{\|\V{\sigma}_1\|\rightarrow +\infty} \cL_1(\V{m}_1,\V{m}_2,\V{\Sigma}_1,K_0;\V{Z}) &= -\infty,
\\
\lim_{\|\V{\sigma}_1\|\rightarrow 0} \cL_1(\V{m}_1,\V{m}_2,\V{\Sigma}_1,K_0;\V{Z}) &= -\infty.
\end{align}
Therefore, searching the stationary points of $\cL_1(\V{m}_1,\V{m}_2,\V{\Sigma}_1,K_0;\V{Z})$ in the interior of its 
domain leads to\footnote{The sign of the derivative allows to establish that the solution represents a maximum point.}
\be
\widehat{\sigma}^2_{1,n}=\frac{1}{K}\left[ \sum_{k=1}^{K_0} (z_{k,n}-\widehat{m}_{1,n})^2  + \sum_{k=K_0+1}^{K} (z_{k,n}-\widehat{m}_{2,n})^2\right]
\ee
and the final decision rule is given by
\begin{multline}
\dmax_{K_0} \Bigg\{\frac{K}{2} \sum_{n=1}^N \log \left[
\frac{1}{K}\sum_{k=1}^K (z_{k,n}-\widehat{m}_{0,n})^2
\right]
\\
-\frac{K}{2}
\sum_{n=1}^N \log 
\Bigg[
\frac{1}{K}
\Bigg(\sum_{k=1}^{K_0} (z_{k,n}-\widehat{m}_{1,n})^2
\\
+\sum_{k=K_0+1}^{K} (z_{k,n}-\widehat{m}_{2,n})^2
\Bigg)
\Bigg]\Bigg\}\test\eta.
\end{multline}
In the next sections, we will refer to this architecture as SPoofing Detector for Uncorrelated Measurements (SP-D-UM).

\subsection{GLRT for Correlated Measurements}
In this case, the covariance matrix $\V{\Sigma}_i$, $i=0,1$, exhibits a general symmetric structure. Thus, under $H_0$, 
the \ac{MLE}s of $\V{m}_0$ and $\V{\Sigma}_0$ are given by $\bar{\V{m}}_0$ and 
$\bar{\V{\Sigma}}_0$, whereas, under $H_1$, it is not difficult to show that 
%\be
$\cL_1(\V{m}_1,\V{m}_2,\V{\Sigma}_1,K_0;\V{Z})\leq \cL_1(\V{m}_1,\V{m}_2,\bar{\V{\Sigma}}_1,K_0;\V{Z})$
%\ee
with
\begin{multline}
\bar{\V{\Sigma}}_1(\V{m}_1,\V{m}_2) = \frac{1}{K}\Bigg[\sum_{k=1}^{K_0} (\V{z}_k-\V{m}_1)(\V{z}_k-\V{m}_1)^T
\\
+\sum_{k=K_0+1}^{K} (\V{z}_k-\V{m}_2)(\V{z}_k-\V{m}_2)^T\Bigg].
\end{multline}
Now, approximate the compressed log-likelihood as
\begin{align}
&\cL_1(\V{m}_1,\V{m}_2,\bar{\V{\Sigma}}_1,K_0;\V{Z}) \nonumber
\\
&\approx -\frac{K}{2}\log\det\Bigg\{\Bigg[\sum_{k=1}^{K_0} (\V{z}_k-\V{m}_1)(\V{z}_k-\V{m}_1)^T \nonumber
\\
&+\sum_{k=K_0+1}^{K} (\V{z}_k-\V{m}_2)(\V{z}_k-\V{m}_2)^T\Bigg] \Bigg\} \nonumber
\\
&=-\frac{K}{2}\log\det\{ \V{M}_0+\V{M}_1+\V{u}_0\V{u}_0^T+\V{u}_1\V{u}_1^T\} \nonumber
\\
& \leq -\frac{K}{2}\log\det\{ \V{M}_0+\V{M}_1\},
\end{align}
where the inequality is due to the fact that $\V{u}_0\V{u}_0^T+\V{u}_1\V{u}_1^T$ is positive semidefinite \cite{MatrixAnalysis}; 
the equality holds when $\V{u}_0=\V{u}_1=0$, namely
\be
\bar{\V{m}}_{1} = \ds\frac{1}{K_0}\sum_{k=1}^{K_0} \V{z}_{k}, \qquad \bar{\V{m}}_{2} = \ds\frac{1}{K_1}\sum_{k=K_0+1}^{K} \V{z}_{k}. 
\label{eq:MLE_m_Spoof_Corr}
\ee
Finally, the logarithm of the GLRT has the following form
\begin{multline}
\dmax_{K_0}\Bigg\{-\log\det\Bigg[ \ds\sum_{k=1}^{K_0} (\V{z}_k-\bar{\V{m}}_1)(\V{z}_k-\bar{\V{m}}_1)^T
\\
 +\sum_{k=K_0+1}^{K} (\V{z}_k-\bar{\V{m}}_2)(\V{z}_k-\bar{\V{m}}_2)^T \Bigg]\Bigg\}
\\
+\log\det \left[ \sum_{k=1}^{K}  (\V{z}_k-\bar{\V{m}}_0)(\V{z}_k-\bar{\V{m}}_0)^T \right]\test \eta.
\end{multline}
The above decision scheme is referred to in the following as SPoofing Detector for Correlated Measurements (SP-D-CM).

\subsubsection{\ac{LVM} for Correlated Measurements}
The $K$ discrete random variables $\omega_k$s introduced in Subsection \ref{subsec:LVM_corr_NLJ} now are such that 
when $\omega_k=a$, $a\in\cA$, then $\V{z}_k\sim\cN_N(\V{m}_a,\V{\Sigma})$ under $H_1$.
Therefore, the \ac{PDF} of $\V{z}_k$ can be written exploiting the Total Probability Theorem as
\be
f_1(\V{z}_k;\V{\pi},\V{m}_1,\V{m}_2,\V{\Sigma}_1)=\sum_{a\in\cA} \pi_a f_1(\V{z}_k;\V{m}_a,\V{\Sigma}).
\label{eq:pdf_LVM}
\ee
The estimates of the unknown parameters are obtained by applying the \ac{EM} algorithm and suitably leveraging the results of
Subsection \ref{subsec:LVM_corr_NLJ}. Specifically, the E-step is accomplished by noticing that 
\begin{multline}
\log f_1(\V{Z};\V{\pi},\V{m}_1,\V{m}_2,\V{\Sigma}_1)
\\
\geq \sum_{k=1}^K \sum_{a\in\cA} \log\left[  \frac{ \pi_a f_1(\V{z}_k;\V{m}_a,\V{\Sigma}_1)}{q_k(a)} \right] q_k(a),
\end{multline}
where the equality holds if
\be
q_k(a)=\frac{ \pi_a f_1(\V{z}_k;\V{m}_a,\V{\Sigma}_1)}{\ds\sum_{m\in\cA} \pi_m f_1(\V{z}_k;\V{m}_a,\V{\Sigma}_1)}.
\ee
It follows that
\be
\tilde{q}^{(h-1)}_k(a)=\frac{ \tilde{{\pi}}_a^{(h-1)} f_1(\V{z}_k;\tilde{\V{m}}_a^{(h-1)},\tilde{\V{\Sigma}}^{(h-1)}_1)}
{\ds\sum_{m\in\cA} \tilde{{\pi}}_m^{(h-1)} f_1(\V{z}_k;\tilde{\V{m}}_a^{(h-1)},\tilde{\V{\Sigma}}^{(h-1)}_1)},
\ee
$\tilde{\V{\pi}}^{(h-1)}$, $\tilde{\V{m}}_1^{(h-1)}$, $\tilde{\V{m}}_2^{(h-1)}$, 
and $\tilde{\V{\Sigma}}_1^{(h-1)}$ are the available estimates at the $(h-1)$th step.

As for the M-step, it leads to the following problem
\begin{align}
&\dmax_{\V{\pi}}\dmax_{\V{m}_1}\dmax_{\V{m}_2}\dmax_{\V{\Sigma}_1}\Bigg\{
\sum_{k=1}^K \sum_{a\in\cA} \tilde{q}^{(h-1)}_k(a)\log(  { f_1(\V{z}_k;\V{m}_a,\V{\Sigma}_1)}) \nonumber
\\
&+\sum_{k=1}^K \sum_{a\in\cA} \tilde{q}^{(h-1)}_k(a)\log(\pi_a) \Bigg\}.
\end{align}
The maximization over $\V{\pi}$ is the same as in Subsection \ref{subsec:LVM_corr_NLJ} and the estimate update is given by
\eqref{eq:priorEstSpoof}. In order to complete the M-step, we have to solve
\be
\dmax_{\V{m}_a \atop a\in\cA}\dmax_{\V{\Sigma}_1}
\sum_{k=1}^K \sum_{a\in\cA} \tilde{q}^{(h-1)}_k(a)\log(  { f_1(\V{z}_k;\V{m}_a,\V{\Sigma}_1)}).
\label{eq:LVM_Mstep}
\ee
To this end, notice that the objective function can be recast as
\begin{multline}
\frac{K}{2}\Bigg\{
\log\det[\V{\Sigma}_1^{-1}]-\tr\Bigg[\V{\Sigma}_1^{-1} 
\\
\times \frac{1}{K} \sum_{k=1}^K\sum_{a\in\cA} 
\tilde{q}^{(h-1)}_k(a) (\V{z}_k-\V{m}_a)(\V{z}_k-\V{m}_a)^T
\Bigg]\Bigg\},
\end{multline}
and, hence, the maximum with respect to $\V{\Sigma}_1$ is attained at
\be
\tilde{\V{\Sigma}}_1^{(h)}=\frac{1}{K} \sum_{k=1}^K\sum_{a\in\cA} 
\tilde{q}^{(h-1)}_k(a) (\V{z}_k-\V{m}_a)(\V{z}_k-\V{m}_a)^T.
\ee
Therefore, the last optimization problem is given by
\begin{multline}
\dmax_{\V{m}_a \atop a\in\cA}\Bigg\{
-\frac{K}{2}\log\det\Bigg[
\sum_{k=1}^K\sum_{a\in\cA} 
\tilde{q}^{(h-1)}_k(a) 
\\
\times(\V{z}_k-\V{m}_a)(\V{z}_k-\V{m}_a)^T\Bigg]\Bigg\},
\label{eq:LVM_Mstep}
\end{multline}
which can be solved by exploiting the procedure used to obtain \eqref{eq:MLE_m_Spoof_Corr}.
As a consequence, the maximum is attained at
\be
\tilde{\V{m}}_a^{(h)}=\frac{1}{q^{(h-1)}(a)}\sum_{k=1}^K\tilde{q}^{(h-1)}_k(a)\V{z}_k.
\ee
Gathering the above results, we can write the modified GLRT as
\be
\label{eq:LVMSpoof}
\frac{\ds\prod_{k=1}^K\sum_{a\in\cA} \tilde{\pi}_a^{(h)} f_1(\V{z}_k;\tilde{\V{m}}_a^{(h)},\tilde{\V{\Sigma}}_1^{(h)})}
{\ds f_0(\V{Z};\bar{\V{m}}_0,\bar{\V{\Sigma}}_0)}\test \eta.
\ee
Finally, we will refer to this decision rule as LVM-based SPoofing Detector (LVM-SP-D).

\section{Illustrative Examples and Discussion}
\label{sec:performanceAnalysis}
In this section, we present a case study to evaluate the performance of the proposed detectors 
assuming that the localization function is based on 
ranging, \ac{DOA} estimation, and \ac{RSRP} measurements. As for the performance metrics, 
we adopt the probability of detection ($P_\text{d}$) as a function of the parameter variation extent and for a preassigned value
of the $P_\text{fa}$. We first proceed with the description of the general 
simulation setting  and, then,  show the detection performance of the proposed architectures for both scenarios where either \ac{NLJ} or spoofing attacks take place. 

\subsection{Simulation Settings}
Consider a scenario with a \ac{UE} that is localized based upon range and \ac{DOA} estimates (azimuth and elevation angles) 
from a single \ac{AN}. As a consequence, the vector size in \eqref{eq:meas} is $N=3$. 
The nominal distance between the \ac{AN} and the \ac{UE} is $200\,$m with an \ac{SNR} equal to $-20\,$dB.
Since the choice of the measurement error model is out of the scope of this paper, 
we borrow exemplary models from existing literature. For example, we model the range error 
as resulted in \cite{PerRenGenRauDomFerBlaCueChaBarGreRiePriLopSec:19} where ranging is 
performed through  \ac{DL}-\ac{TDOA} measurements of 5G positioning reference signal in 
a urban macro environment with line-of-sight conditions. 
The \ac{DOA} measurements are characterized as in \cite{RasCosKoiLepVal:18} where the angle estimates are 
obtained through a beam-\ac{RSRP} of \ac{DL} with $16$ \ac{UE} beams.

Since deriving closed-form expressions for the $P_\text{d}$ and $P_\text{fa}$ is not an easy task at least to the best of authors' knowledge,
we resort to standard Monte Carlo counting techniques where the $P_\text{d}$ and the detection thresholds are estimated
over $1000$ and $100/P_\text{fa}$ independent trials, respectively, with $P_\text{fa}=10^{-2}$.
The proposed decision schemes are assessed in scenarios accounting for two different lengths of the sliding window. Specifically,
we consider $K=24,32$. As for the value of $K_0$, we assume that $K_0 \in \{K/4, K/2, 3K/4 \}$.

The \ac{NLJ} attack is simulated by varying the variance of the noise affecting the measurements. Specifically, 
starting from a diagonal covariance matrix, $\M{\Sigma}_0$ say, set using the results of the aforementioned references, 
we modify the latter through a scaling factor $\gamma$ such that 
$\M{\Sigma}_1= \M{\Sigma}_0$ and $\M{\Sigma}_2=  \gamma \M{\Sigma}_0$ in \eqref{eq:hyptestNLJ}. As a case study, we consider the 
true range and \ac{DOA} (elevation and azimuth) when the distance between the \ac{AN} and UE is 
equal to $d_\text{0}=200\,$m and the DOA is 0 degrees. Then, the value of $\V{m}_0=\V{m}_1$ is set by applying to the true range and DOA values an error modeled according to the exemplary error \acp{PDF} proposed in \cite{PerRenGenRauDomFerBlaCueChaBarGreRiePriLopSec:19} for the TDOA and \cite{RasCosKoiLepVal:18} for the DOA, with a SNR of $-20\,$dB. 
On the other hand, in the case of spoofing attack, the numerical examples are obtained by varying the mean value of the original signal.
More precisely, the mean value of the measurements under $H_0$ is multiplied by a factor $\nu$ such that 
$\V{m}_1= \V{m}_0$ and $\V{m}_2=  \nu \V{m}_0$  in \eqref{eq:hyptestSpoof}. In this case, the covariance 
matrix $\V{\Sigma}_0=\V{\Sigma}_1$ is obtained as the empirical covariance matrix computed over $10^4$ TDOA and 
DOA measurements generated using the exemplary error \acp{PDF} proposed 
in \cite{PerRenGenRauDomFerBlaCueChaBarGreRiePriLopSec:19} for the TDOA and in \cite{RasCosKoiLepVal:18} 
for the DOA, with a SNR of $-20\,$dB. 
 
%{\color{red} \bf qui dobbiamo dire come impostiamo }

Before concluding this section, two important remarks are in order. First, we set
the number of iterations for the EM-based architectures to $10$. This choice ensures 
a reasonable trade off between
computational burden and convergence issues as shown in Figure \ref{fig:convergenceLVM}, where
we show the \ac{RMS} values for
\be
\Delta(\V{Z},h)=\left|\frac{t^{(h)}(\V{Z})-t^{(h-1)}(\V{Z})}
{t^{(h)}(\V{Z})}\right|
\ee
with $t^{(h)}(\V{Z})$ the numerator of either \eqref{eq:NLJLVM} or \eqref{eq:LVMSpoof}, versus $h$.
Second, notice that the $P_\text{fa}$ of the considered
architectures is not very sensitive
to variations of the parameter values used to set the detection thresholds as shown in Figure \ref{fig:PFAvsMeant}, where we plot 
the $P_\text{fa}$ for the different attacks and 
proposed techniques as a function of the unknown parameters under $H_0$. 
In particular, the threshold is fixed targeting a false alarm probability $P_\text{fa}^{\star}=10^{-2}$ and 
considering $\V{m}^{\prime}_0$ 
and $\M{\Sigma}^{\prime}_0$ as mean and covariance under $H_0$, with $K=32$.
Figure \ref{fig:PFAvsMean1} contains the $P_\text{fa}$ for $\V{m}^{\prime}_0=\nu\V{m}_0$ 
versus $\nu$ and highlights that all the curves belong to the interval $[0.009, \ 0.011]$.
The sensitivity to the covariance matrix is shown in Figure \ref{fig:PFAvsVar1} where the $P_\text{fa}$ 
is computed when $\M{\Sigma}^{\prime}_0=\gamma \M{\Sigma}_0$. 
In this case, the LVM-SP-D is not capable of maintaining the value of the $P_\text{fa}$ within the same interval as the other decision rules.
For this reason, from a practical point of view, this architecture would require a continuous monitoring of the noise power level
in order to select the right threshold.

\begin{figure}[tbp]
\centering
\includegraphics[width=0.8\columnwidth]{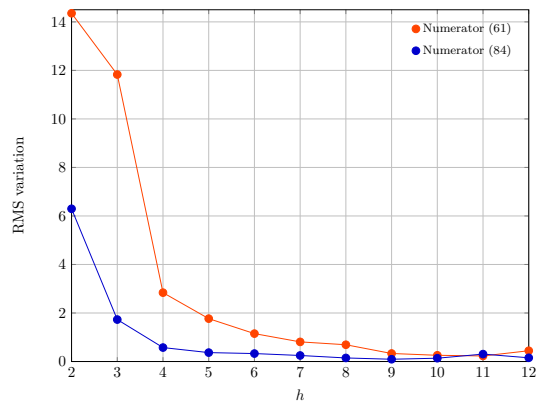}
\caption{RMS values of $\Delta(\V{Z},h)$ versus $h$.}
\label{fig:convergenceLVM}
\end{figure}

\begin{figure*}[!tb]
\centering
\begin{subfigure}[][$\V{m}^{\prime}_0=\nu\V{m}_0$ ]{\label{fig:PFAvsMean1}
\includegraphics[width=0.8\columnwidth]{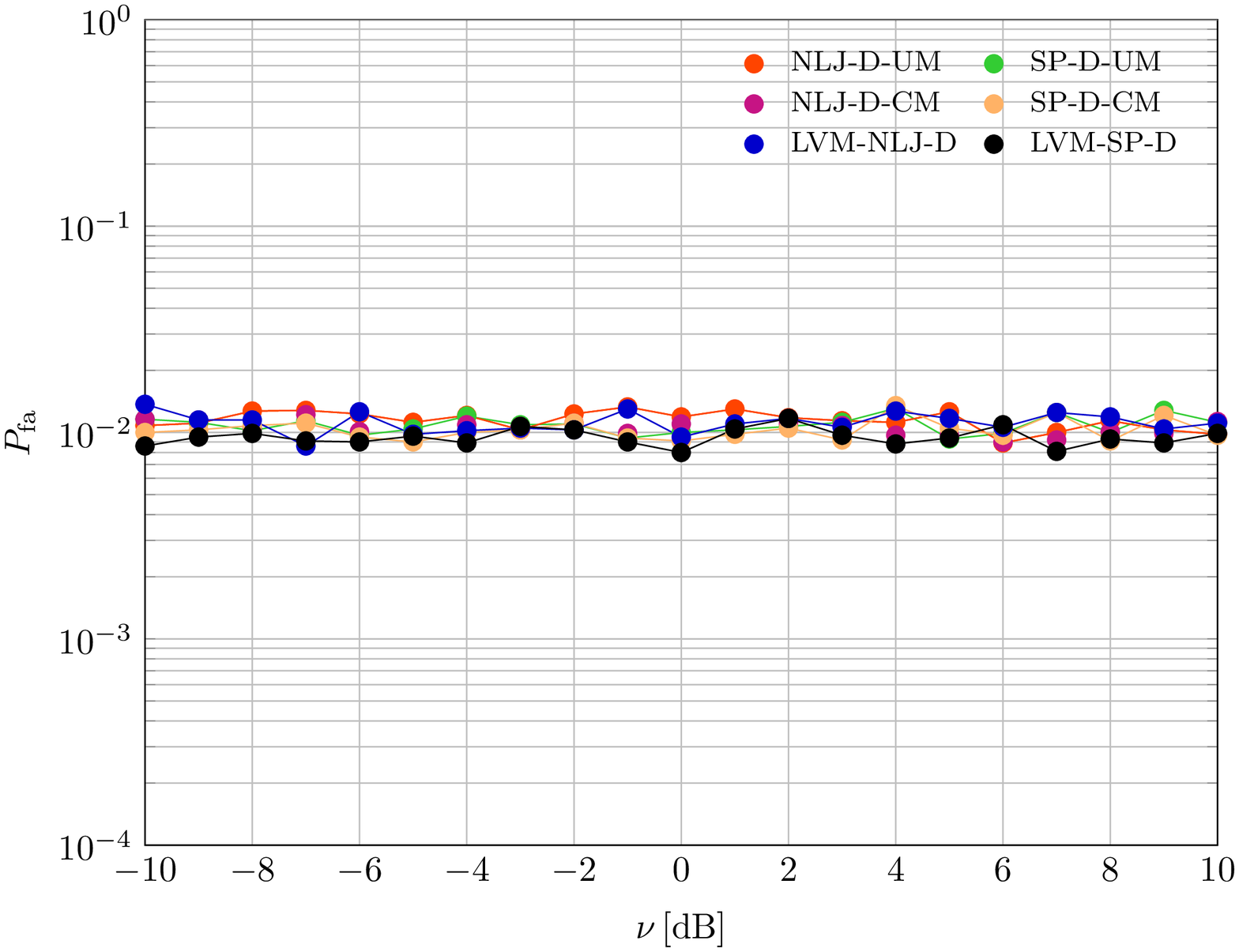}
}
\end{subfigure}
\begin{subfigure}[][$\V{\Sigma}^{\prime}_0=\gamma\V{\Sigma}_0$]{
\label{fig:PFAvsVar1}
\includegraphics[width=0.8\columnwidth]{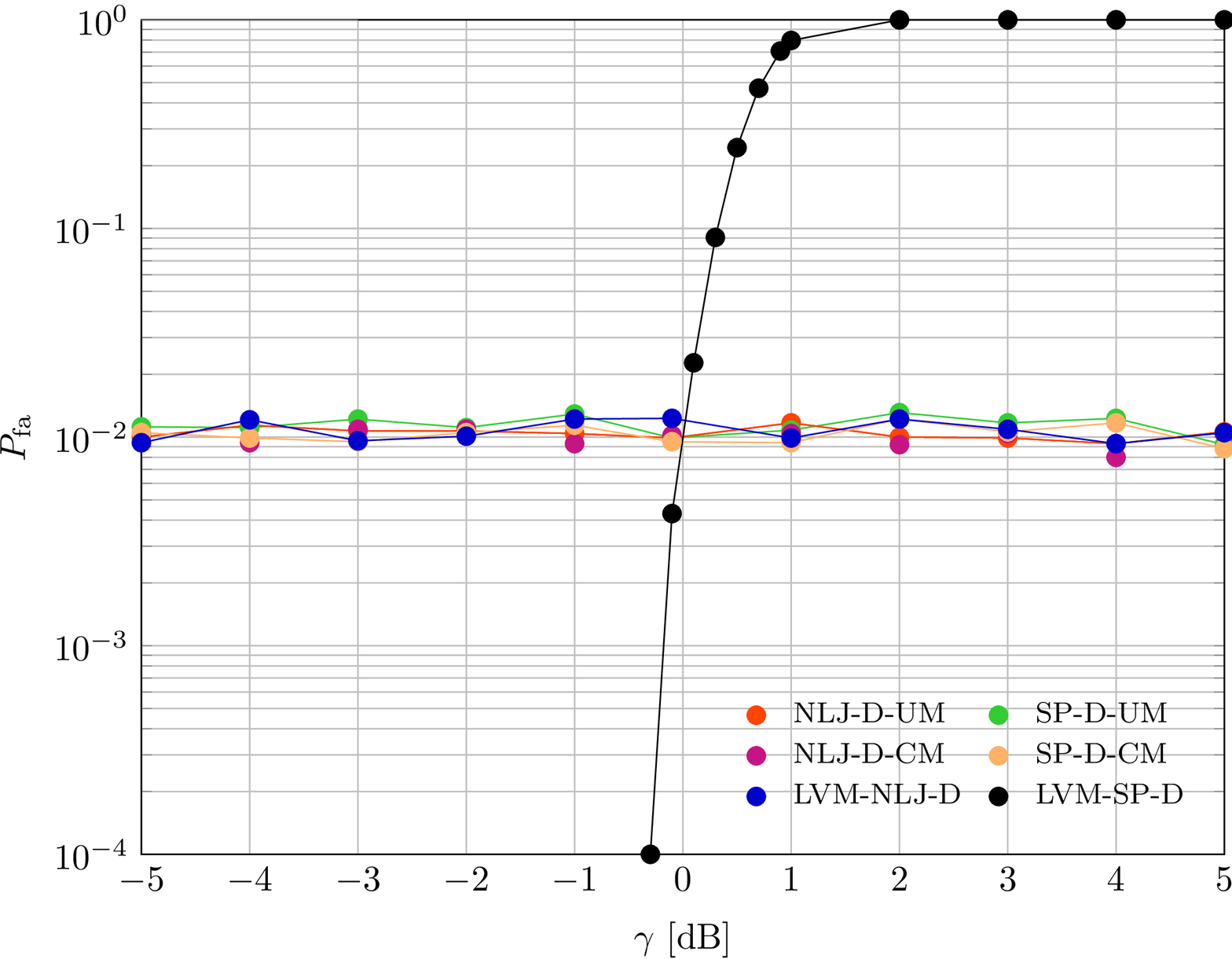}}
\end{subfigure}
\caption{$P_\text{fa}$ versus $\nu$ and $\gamma$. In (a) the $P_\text{fa}$ is computed assuming $\V{m}^{\prime}_0=\nu\V{m}_0$ and is 
equal to the nominal value $10^{-2}$ when $\V{m}^{\prime}_0=\V{m}_0$. In (b)  the $P_\text{fa}$ is computed 
assuming $\V{\Sigma}^{\prime}_0=\gamma\V{\Sigma}_0$ and is equal to the nominal value $10^{-2}$ when $\V{\Sigma}^{\prime}_0=\V{\Sigma}_0$.}
\label{fig:PFAvsMeant}
\end{figure*}

%\begin{figure}[htp]
%\centering
%\includegraphics[width=0.95\columnwidth]{Figures/PfaVsMean.eps}
%\caption{$P_\text{fa}$ versus $\nu$, where the $P_\text{fa}$ is computed assuming $\V{m}^{\prime}_0=\nu\V{m}_0$ and using a
%threshold returning the nominal value $10^{-2}$ when $\V{m}^{\prime}_0=\V{m}_0$.}
%\label{fig:PFAvsMean}
%\end{figure}
%%
%\begin{figure}[htp]
%\centering
%\includegraphics[width=0.95\columnwidth]{Figures/PfaVsVar.eps}
%\caption{$P_\text{fa}$ versus $\gamma$, where the $P_\text{fa}$ is computed assuming $\V{\Sigma}^{\prime}_0=\gamma\V{\Sigma}_0$ and using a
%threshold returning the nominal value $10^{-2}$ when $\V{\Sigma}^{\prime}_0=\V{\Sigma}_0$.}
%\label{fig:PFAvsVar}
%\end{figure}

\subsection{Performance of the NLJ Detection Architectures}
%{\color{red} \bf nelle figure, via il title, poi sulle ordinate mettiamo $P_\text{d}$ e sulle ascisse $\gamma$ (dB) o $\nu$ (dB).}
The detection performances for the NLJ-D-UM, NLJ-D-CM, and LVM-NLJ-D are shown in Figures \ref{fig:NLJ-24} and \ref{fig:NLJ-32}, where we plot
the $P_\text{d}$ curves against the parameter $\gamma$.

%\begin{figure}[htp]
%\centering
%\includegraphics[width=0.7\columnwidth]{Figures/Detection_NLJ_Uncorrelated_24.eps}
%\caption{$P_\text{d}$ versus $\gamma$ for the NLJ-D-UM assuming $K=24$, $\mbox{SNR}=-30$ dB (solid lines), $\mbox{SNR}=-15$ dB (dashed lines), 
%and $\mbox{SNR}=0$ dB (dot-dashed lines).}
%\label{fig:NLJUNC24}
%\end{figure}
%\begin{figure}[htp]
%\centering
%\includegraphics[width=0.7\columnwidth]{Figures/Detection_NLJ_Correlated_24.eps}
%\caption{$P_\text{d}$ versus $\gamma$ for the NLJ-D-CM assuming $K=24$, $\mbox{SNR}=-30$ dB (solid lines), $\mbox{SNR}=-15$ dB (dashed lines), 
%and $\mbox{SNR}=0$ dB (dot-dashed lines).}
%\label{fig:NLJCOR24}
%\end{figure}
%\begin{figure}[htp]
%\centering
%\includegraphics[width=0.7\columnwidth]{Figures/Detection_NLJ_LVM_24.eps}
%\caption{$P_\text{d}$ versus $\gamma$ for the LVM-NLJ-D assuming $K=24$, $\mbox{SNR}=-30$ dB (solid lines), $\mbox{SNR}=-15$ dB (dashed lines), 
%and $\mbox{SNR}=0$ dB (dot-dashed lines).}
%\label{fig:NLJLVM24}
%\end{figure}

\begin{figure*}[!t]
\centering
\begin{subfigure}[][NLJ-D-UM]{\label{fig:NLJUNC24}
\includegraphics[width=0.6\columnwidth]{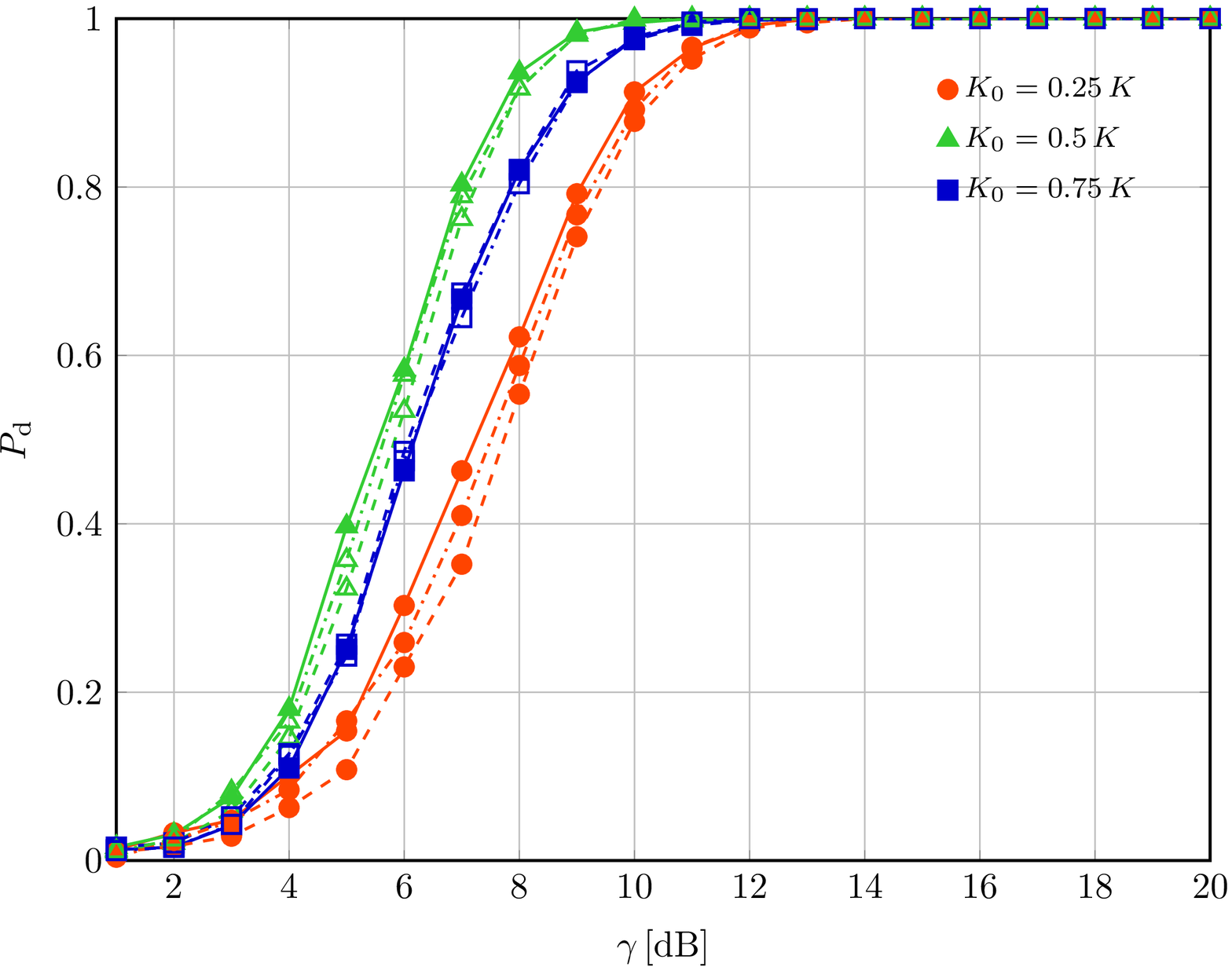}}
\end{subfigure}
\begin{subfigure}[][NLJ-D-CM]{
\label{fig:NLJCOR24}
\includegraphics[width=0.6\columnwidth]{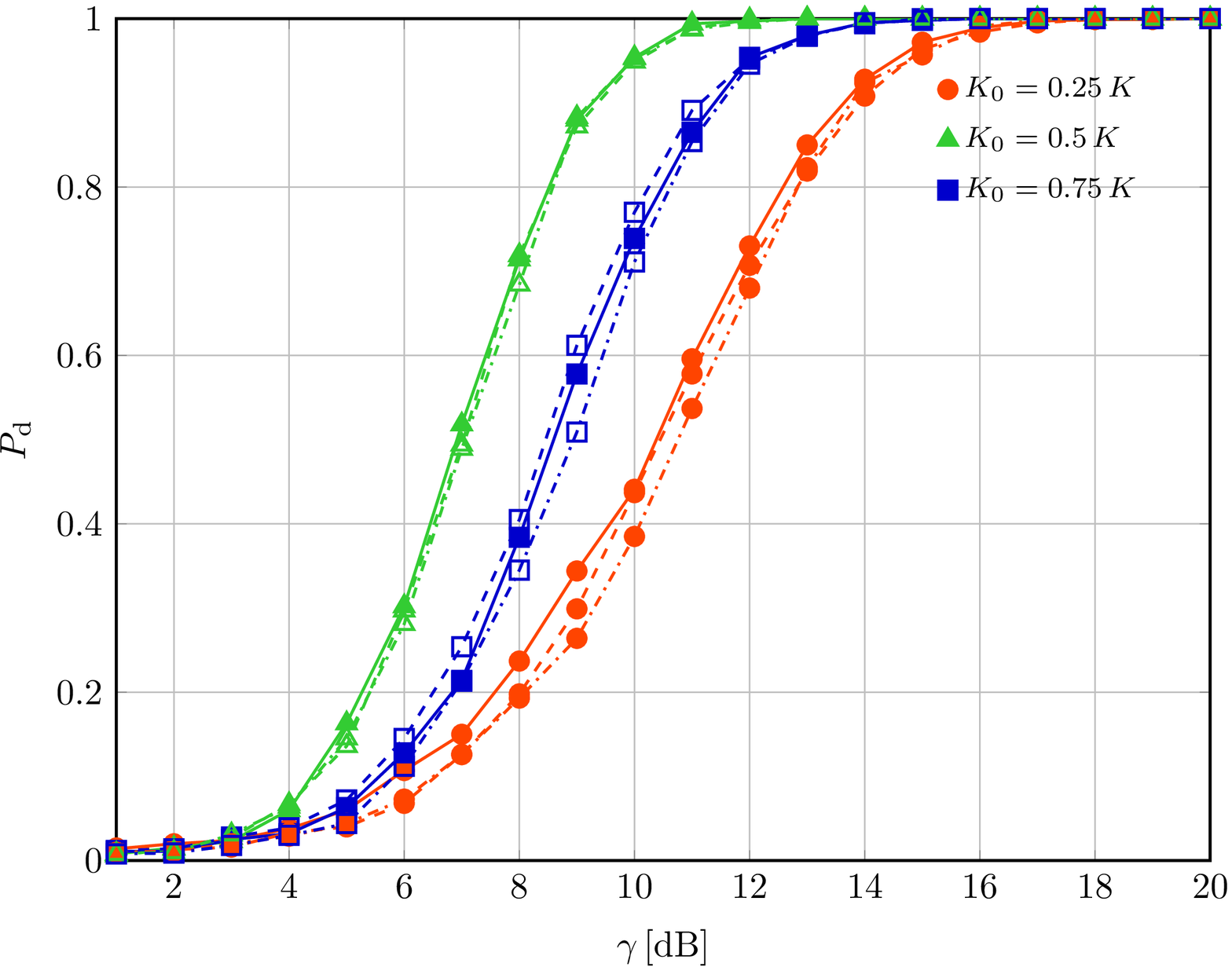}}
\end{subfigure}
\begin{subfigure}[][LVM-NLJ-D]{
\label{fig:NLJLVM24}
\includegraphics[width=0.6\columnwidth]{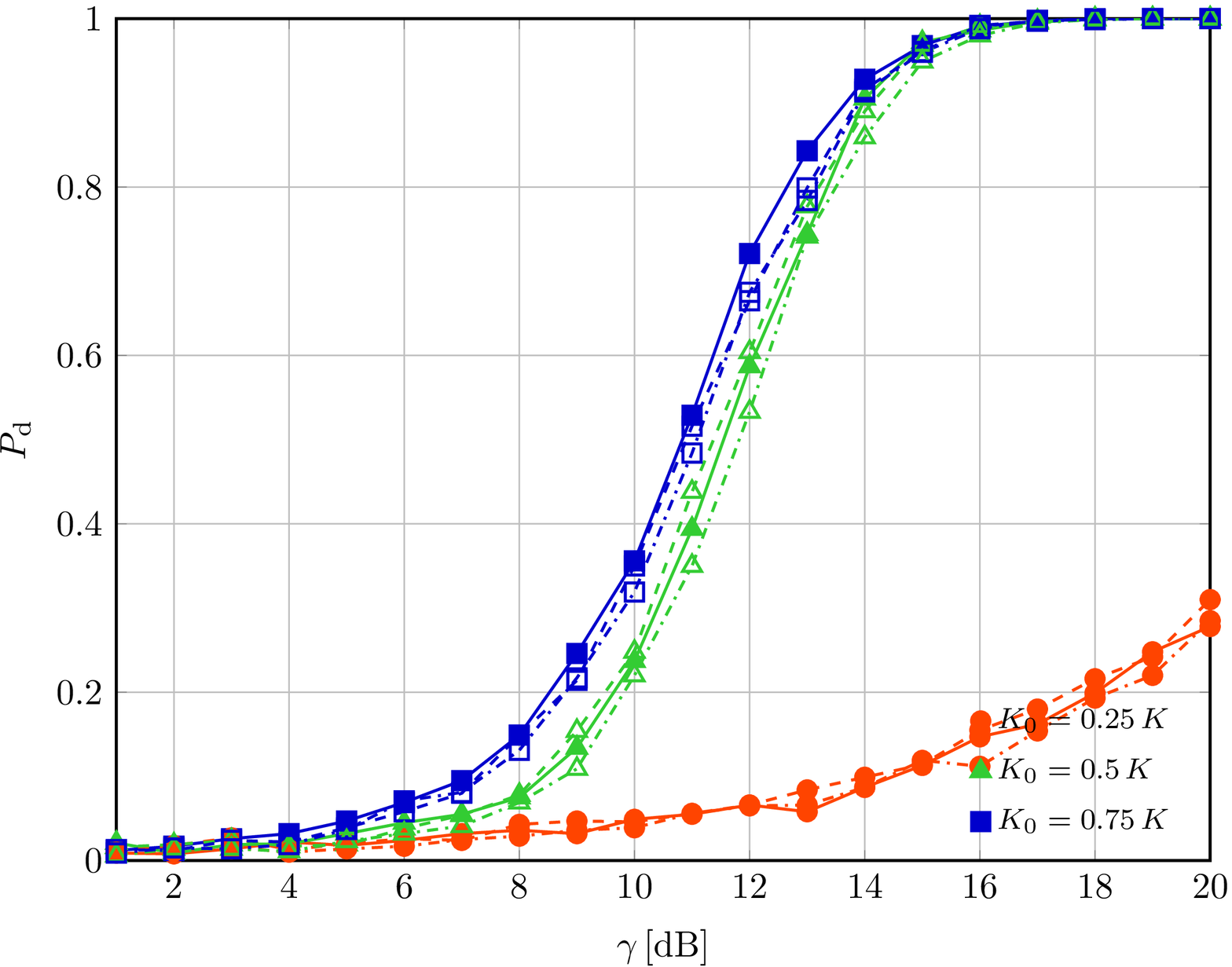}}
\end{subfigure}
\caption{$P_\text{d}$ versus $\gamma$ for the NLJ-D-UM, NLJ-D-CM, and LVM-NLJ-D assuming $K=24$, $\mbox{SNR}=-30$ dB (solid lines), $\mbox{SNR}=-15$ dB (dashed lines), and $\mbox{SNR}=0$ dB (dot-dashed lines).}
\label{fig:NLJ-24}
\end{figure*}

\begin{figure*}[!t]
\centering
\begin{subfigure}[][NLJ-D-UM]{\label{fig:NLJUNC32}
\includegraphics[width=0.6\columnwidth]{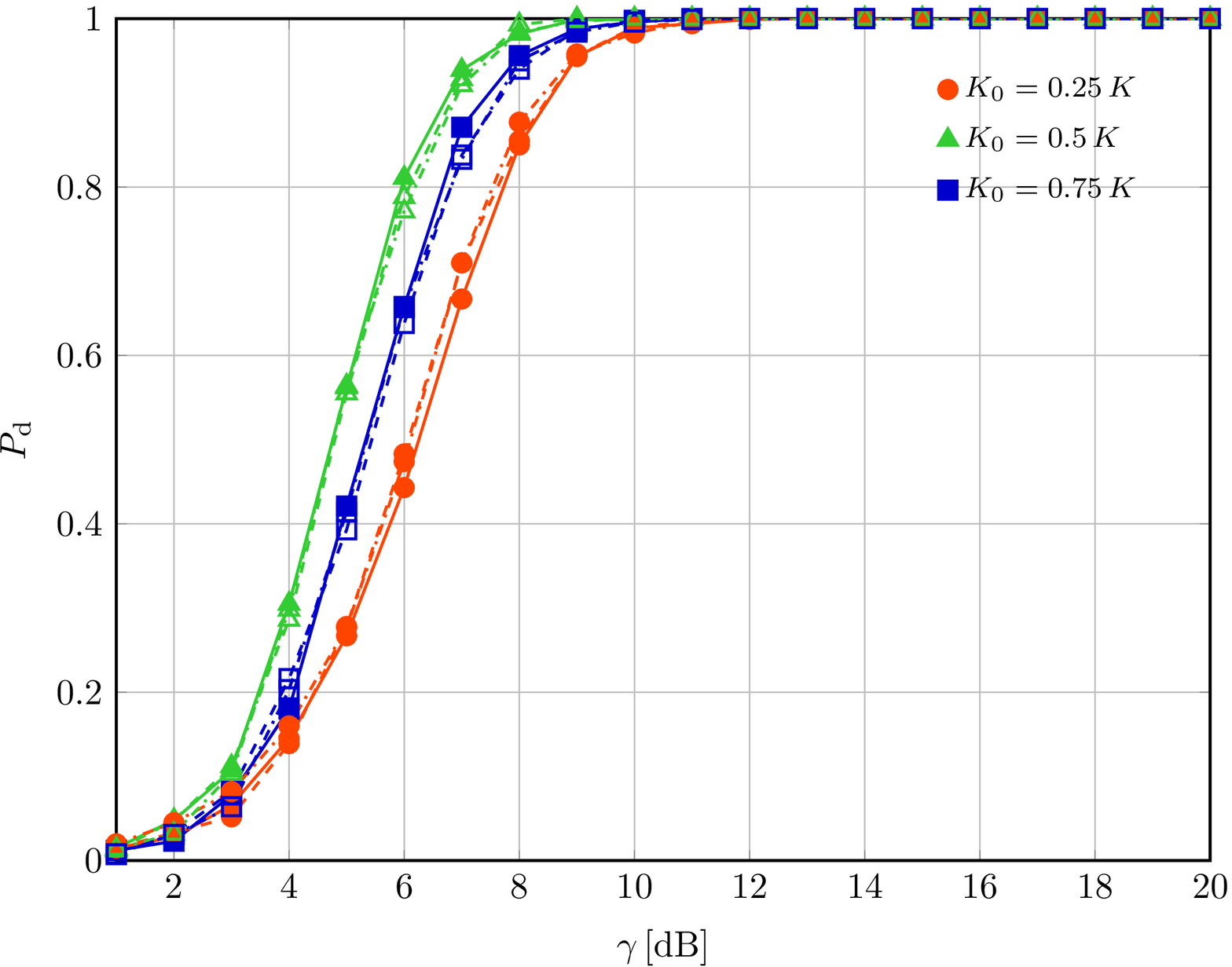}}
\end{subfigure}
\begin{subfigure}[][NLJ-D-CM]{
\label{fig:NLJCOR32}
\includegraphics[width=0.6\columnwidth]{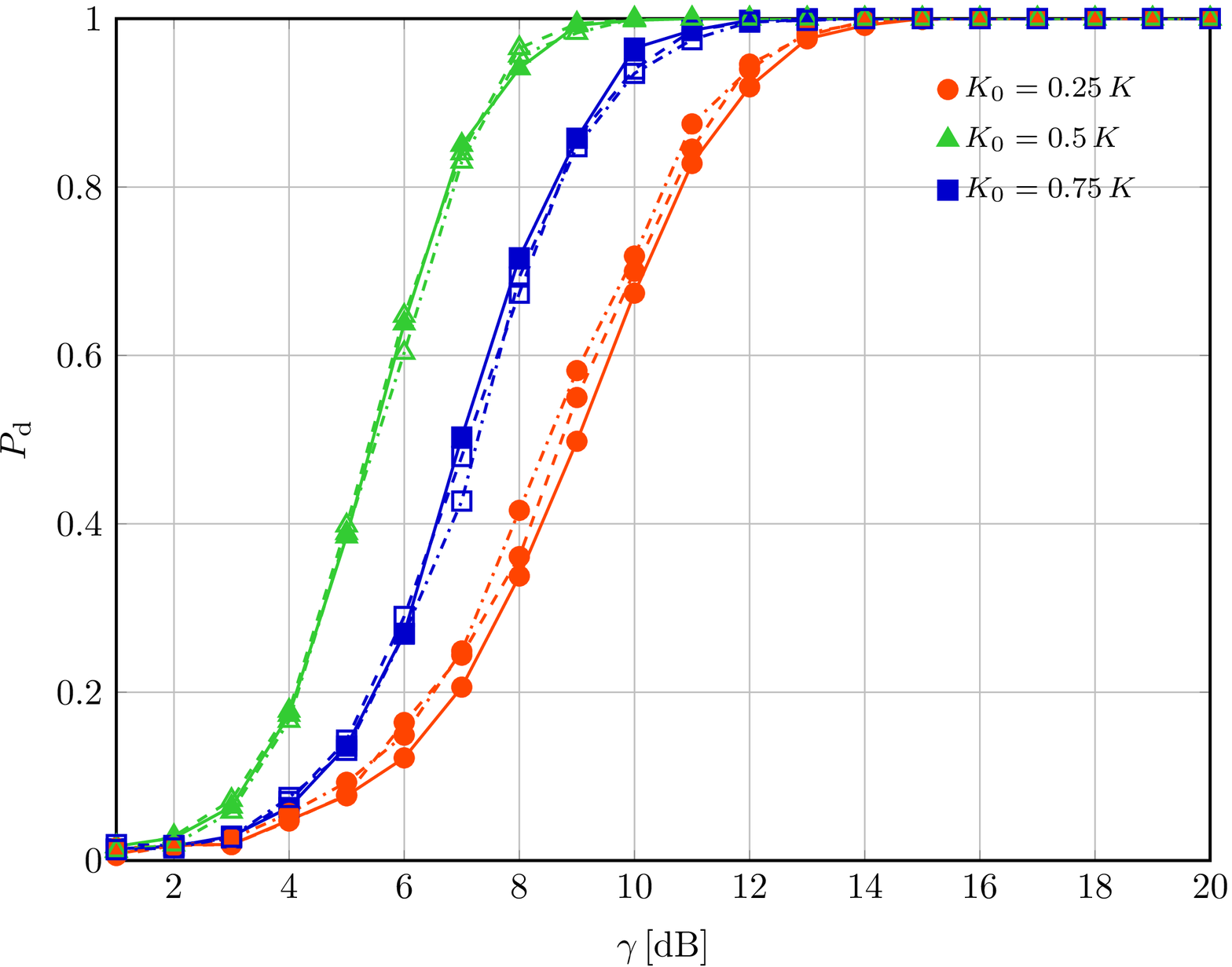}}
\end{subfigure}
\begin{subfigure}[][LVM-NLJ-D]{
\label{fig:NLJLVM32}
\includegraphics[width=0.6\columnwidth]{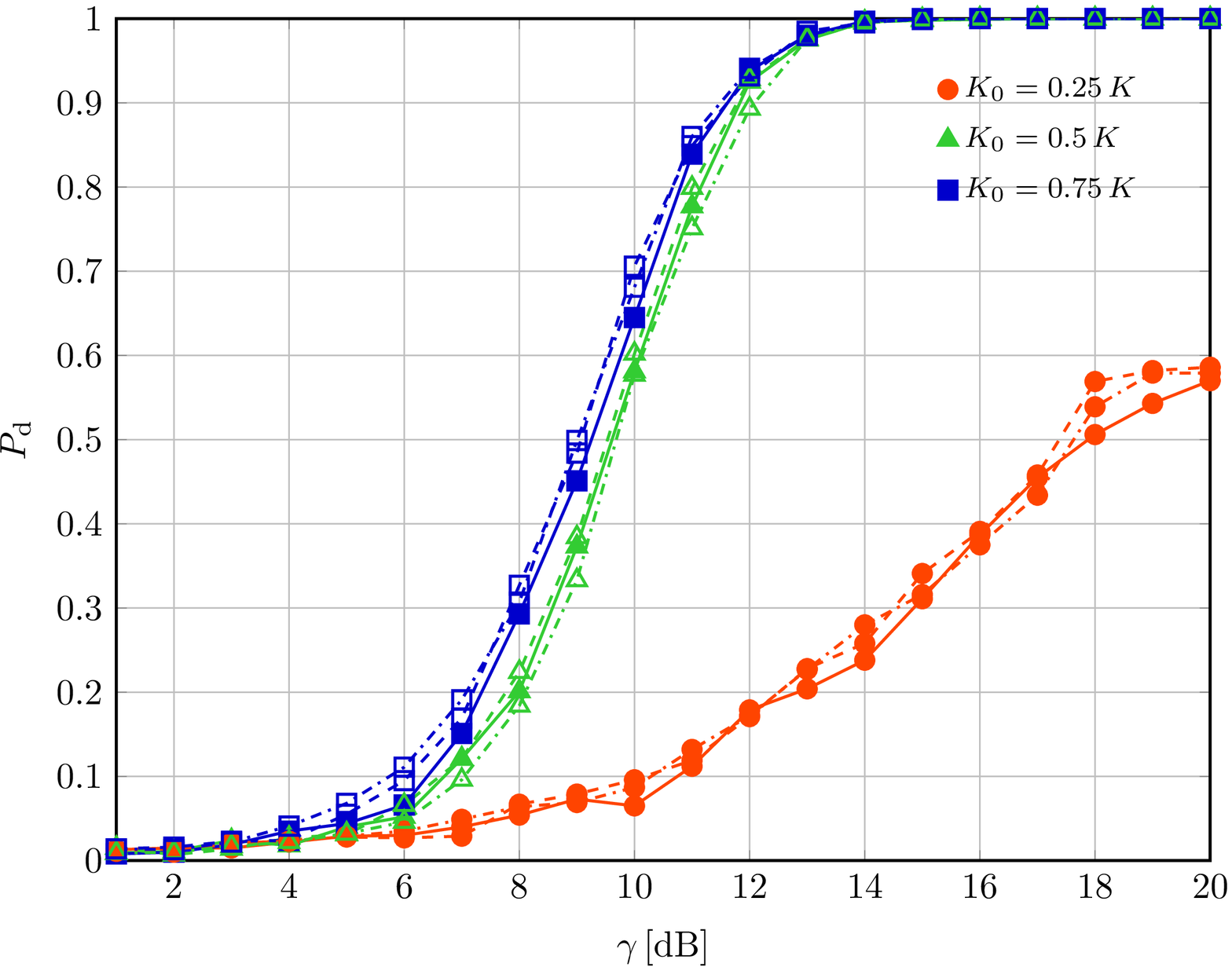}}
\end{subfigure}
\caption{$P_\text{d}$ versus $\gamma$ for the NLJ-D-UM, NLJ-D-CM, and LVM-NLJ-D assuming $K=32$, $\mbox{SNR}=-30$ dB (solid lines), $\mbox{SNR}=-15$ dB (dashed lines), and $\mbox{SNR}=0$ dB (dot-dashed lines).}
\label{fig:NLJ-32}
\end{figure*}

%\begin{figure}[htp]
%\centering
%\includegraphics[width=0.9\columnwidth]{Figures/Detection_NLJ_Uncorrelated_32.eps}
%\caption{$P_\text{d}$ versus $\gamma$ for the NLJ-D-UM assuming $K=32$, $\mbox{SNR}=-30$ dB (solid lines), $\mbox{SNR}=-15$ dB (dashed lines), 
%and $\mbox{SNR}=0$ dB (dot-dashed lines).}
%\label{fig:NLJUNC32}
%\end{figure}
%
%\begin{figure}[htp]
%\centering
%\includegraphics[width=0.9\columnwidth]{Figures/Detection_NLJ_Correlated_32.eps}
%\caption{$P_\text{d}$ versus $\gamma$ for the NLJ-D-CM assuming $K=32$, $\mbox{SNR}=-30$ dB (solid lines), $\mbox{SNR}=-15$ dB (dashed lines), 
%and $\mbox{SNR}=0$ dB (dot-dashed lines).}
%\label{fig:NLJCOR32}
%\end{figure}
%%
%\begin{figure}[htp]
%\centering
%\includegraphics[width=0.9\columnwidth]{Figures/Detection_NLJ_LVM_32.eps}
%\caption{$P_\text{d}$ versus $\gamma$ for the LVM-NLJ-D assuming $K=32$, $\mbox{SNR}=-30$ dB (solid lines), $\mbox{SNR}=-15$ dB (dashed lines), 
%and $\mbox{SNR}=0$ dB (dot-dashed lines).}
%\label{fig:NLJLVM32}
%\end{figure}
Figure \ref{fig:NLJUNC24} shows the performance of the NLJ-D-UM assuming $K=24$ and different values for the \ac{SNR}. 
As expected, while the \ac{SNR} does not have a remarkable impact on the performance, the value of $K_0$ influences the $P_\text{d}$. 
As a matter of fact, the latter does not increase monotonically with $K_0$ and the best performance is obtained 
for $K_0=K/2$. Just to give an example, notice that given $\mbox{\ac{SNR}}=-30\,$dB and $\gamma=8\,$dB, $P_\text{d}=0.61$ 
for $K_0=K/4$, $P_\text{d}=0.92$ for $K_0=K/2$, and $P_\text{d}=0.81$ for $K_0=3K/4$.
The performances of NLJ-D-CM are reported in Figure \ref{fig:NLJCOR24} which shares the same parameter values as in the previous figure.
The results confirm both the insensitivity of the performance to the \ac{SNR} and the previously described behavior with respect
to the value of $K_0$.
For instance, in this case, given \ac{SNR}$=-30\,$dB and $\gamma=8\,$dB, 
we observe $P_\text{d}=0.22$ for $K_0=K/4$, $P_\text{d}=0.71$ for $K_0=K/2$, and $P_\text{d}=0.39$ for $K_0=3K/4$. 
From the comparison between Figures \ref{fig:NLJUNC24} and \ref{fig:NLJCOR24}, it turns out
that NLJ-D-UM exhibits better performance than the NLJ-D-CM since the former exploits the information related to the actual structure of the
error covariance matrix, whereas the latter is designed without assuming any special structure for the covariance matrix.
The performance gain of the NLJ-D-UM over the NLJ-D-CM is considerable for $K_0=K/4$ and $K_0=3K/4$ and
becomes less important for $K_0=K/2$.
Finally, in Figure \ref{fig:NLJLVM24}, we show the $P_\text{d}$ curves associated with the LVM-NLJ-D for $K=24$. The figure points out that this architecture
experiences a very poor performance for the case $K_0=K/4$ and, generally speaking, its 
performance is worse than those of the other detectors.

In Figures \ref{fig:NLJUNC32}-\ref{fig:NLJLVM32}, we analyze the effect of the sliding window size on the detection performance.
In fact, these figures are analogous to the previous three figures except for $K=32$.
As expected, in this case, since the number of available data increases, the estimation quality improves leading to better detection
performance for all the proposed architectures.

As final remark, it is important to underline that despite the fact that the NLJ-D-CM is designed assuming the most general 
structure for the covariance matrix, the detection loss with respect to the NLJ-D-UM, whose performance is estimated
under perfectly matched design conditions, becomes negligible as $K$ increases. Therefore, we would single out the NLJ-D-CM
as an ``all-season'' architecture capable of operating without any a priori assumption about data correlation.

\subsection{Performance of the Spoofing Detection Architectures}
This last subsection focuses on the performance assessment of the SP-D-UM, SP-D-CM, and LVM-SP-D in terms of probability of spoofing detection.
The analysis consider two values of $K$ and three values for the \ac{SNR} as in the previous subsection. Specifically, Figure 
\ref{fig:Spoof24} assume $K=24$ whereas in Figure \ref{fig:Spoof32} $K$ is equal to $32$.

\begin{figure*}[!t]
\centering
\begin{subfigure}[][SP-D-UM]{\label{fig:SpoofUNC24}
\includegraphics[width=0.6\columnwidth]{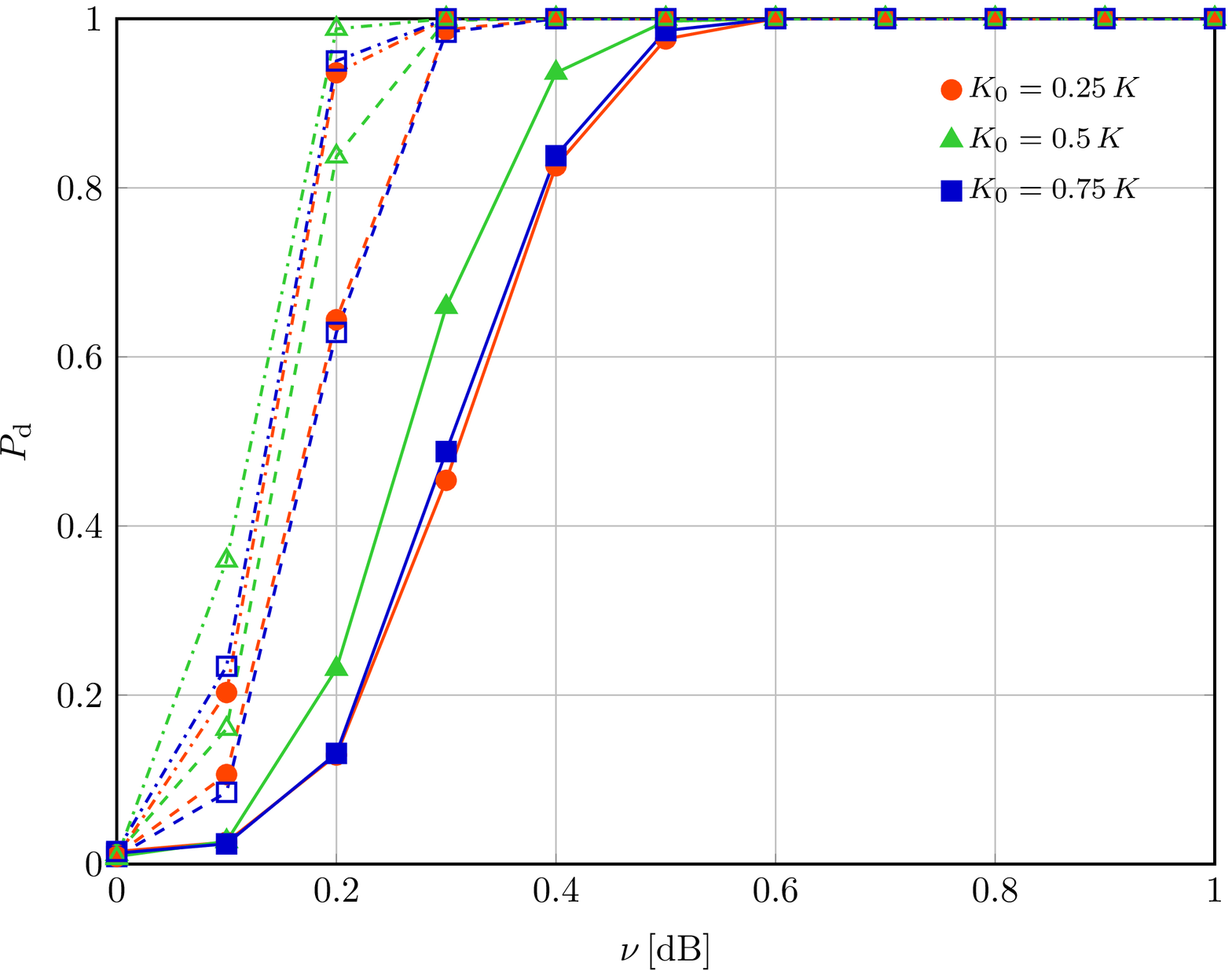}}
\end{subfigure}
\begin{subfigure}[][SP-D-CM]{
\label{fig:SpoofCOR24}
\includegraphics[width=0.6\columnwidth]{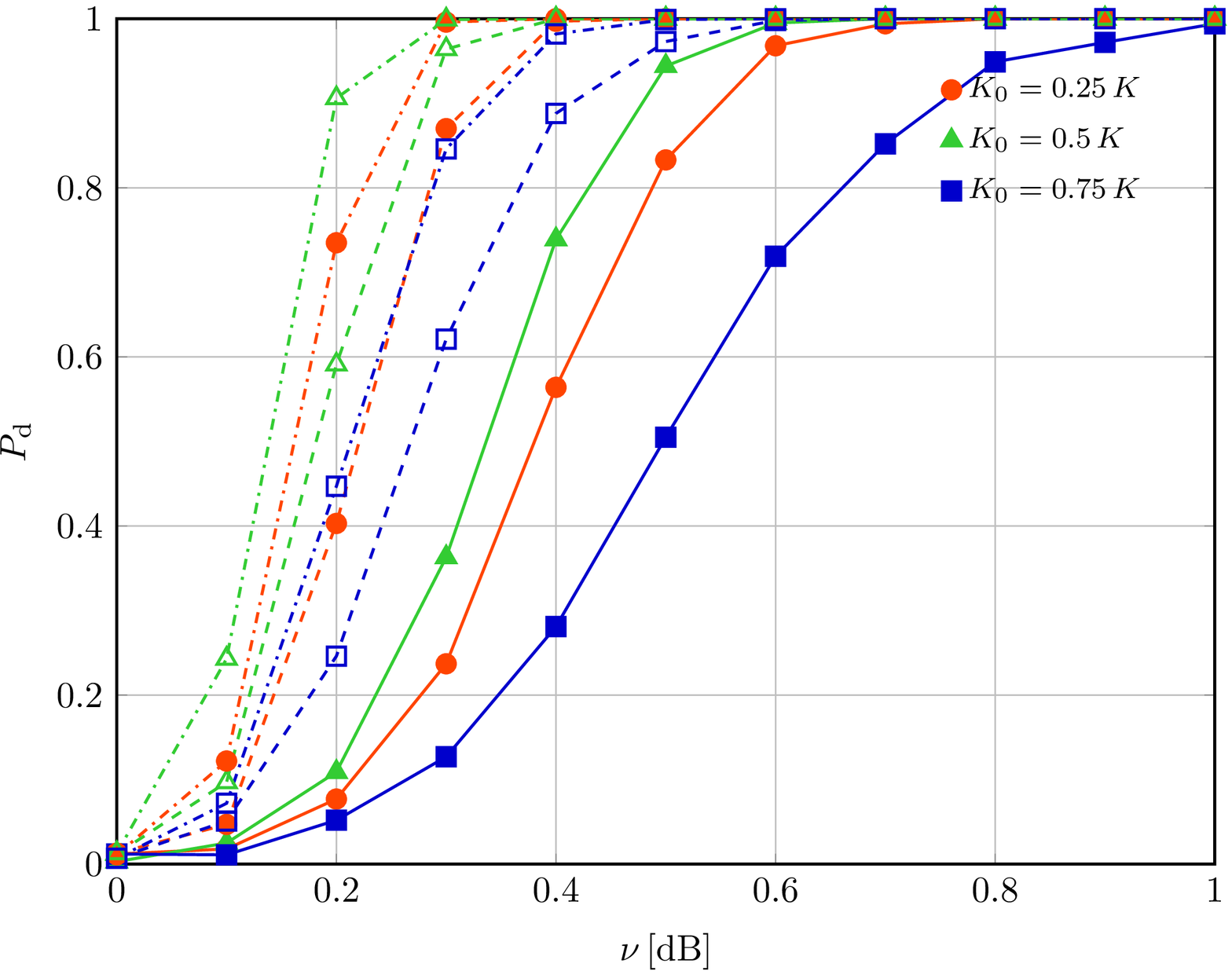}}
\end{subfigure}
\begin{subfigure}[][ LVM-SP-D]{
\label{fig:SpoofLVM24}
\includegraphics[width=0.6\columnwidth]{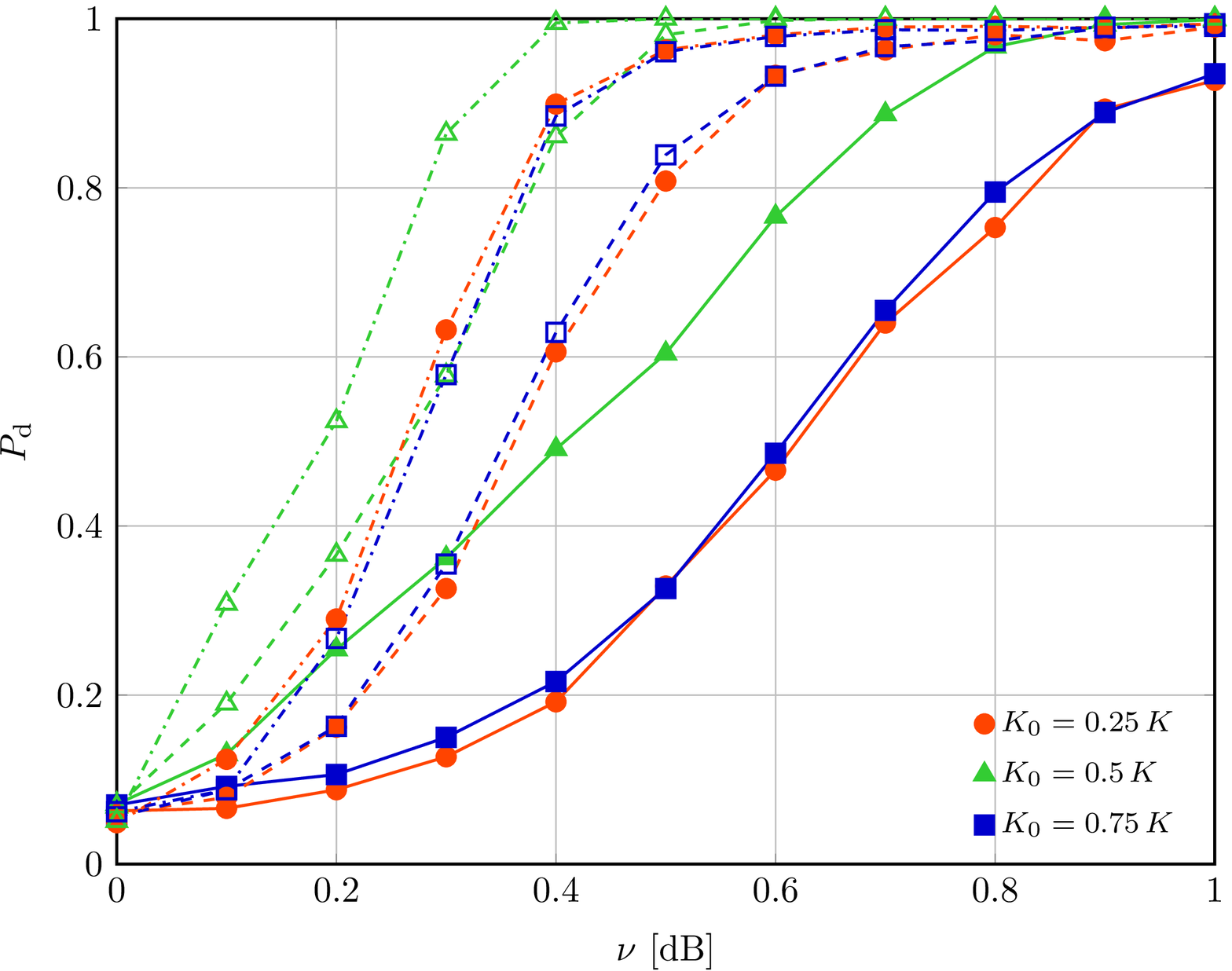}}
\end{subfigure}
\caption{$P_\text{d}$ versus $\nu$ for the SP-D-UM, SP-D-CM, and LVM-SP-D assuming $K=24$, \ac{SNR}$=-30\,$dB (solid lines), \ac{SNR}$=-15\,$dB (dashed lines), and \ac{SNR}$=0\,$dB (dot-dashed lines).}
\label{fig:Spoof24}
\end{figure*}

\begin{figure*}[!t]
\centering
\begin{subfigure}[][SP-D-UM]{\label{fig:SpoofUNC32}
\includegraphics[width=0.6\columnwidth]{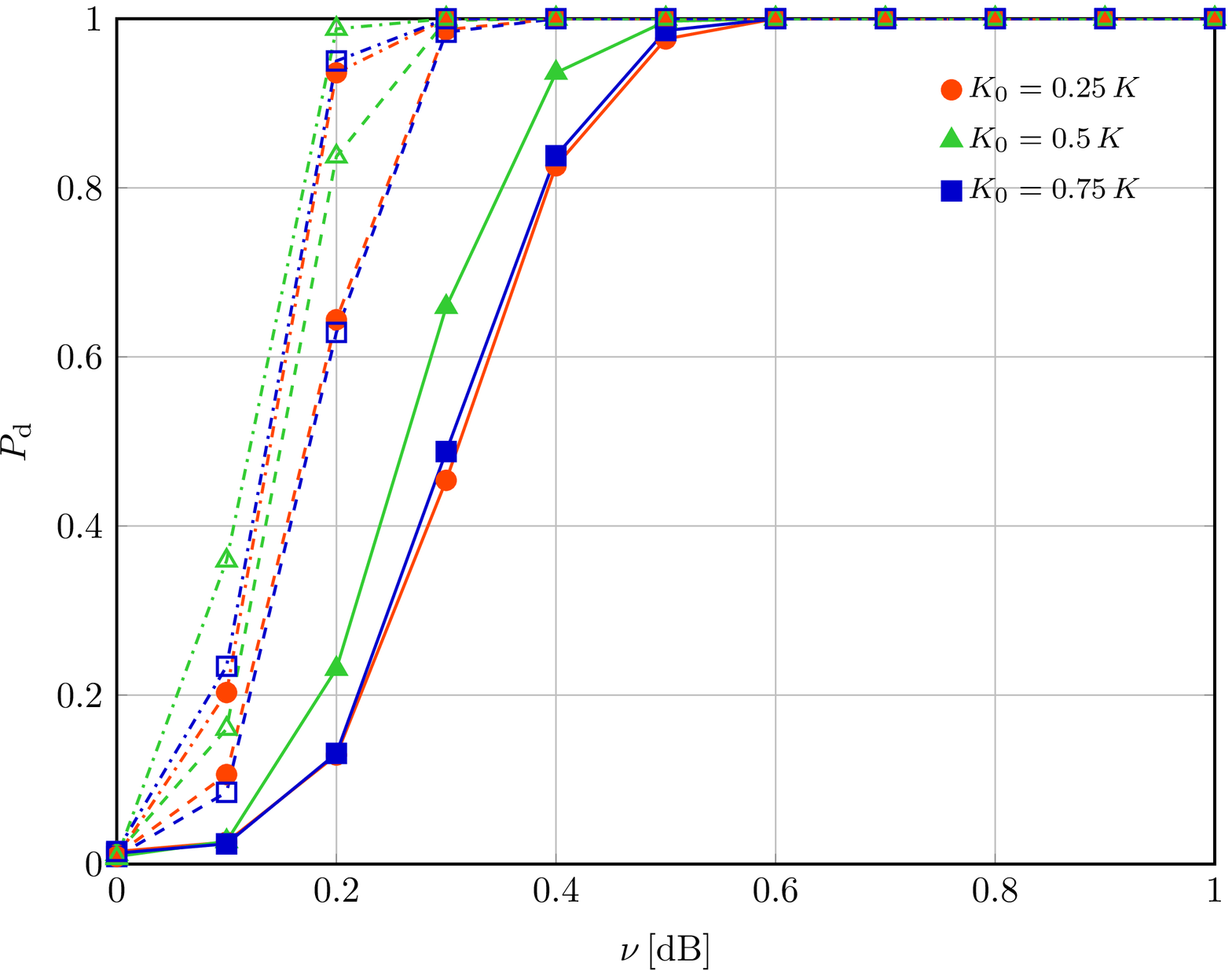}}
\end{subfigure}
\begin{subfigure}[][SP-D-CM]{
\label{fig:SpoofCOR32}
\includegraphics[width=0.6\columnwidth]{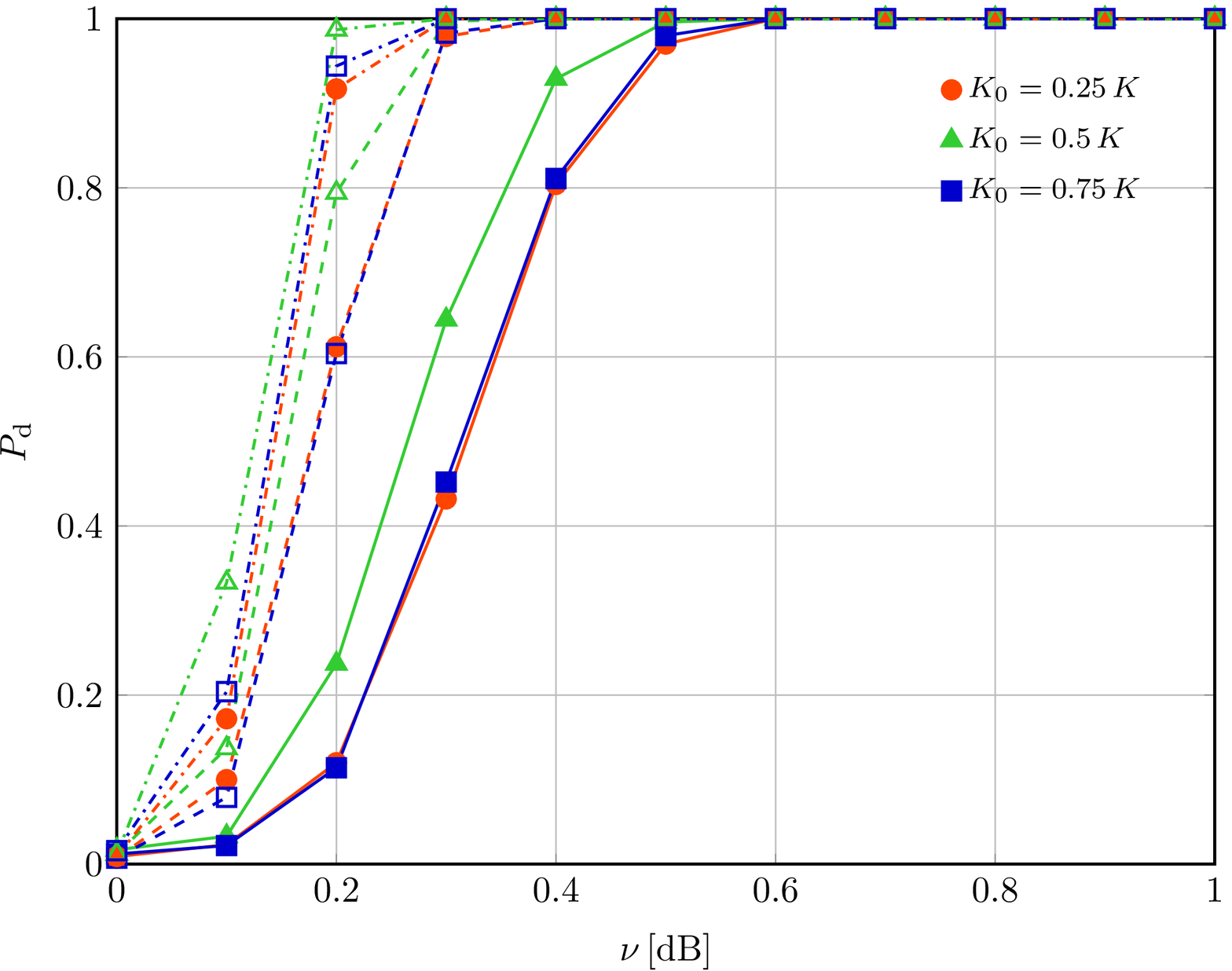}}
\end{subfigure}
\begin{subfigure}[][LVM-SP-D]{
\label{fig:SpoofLVM32}
\includegraphics[width=0.6\columnwidth]{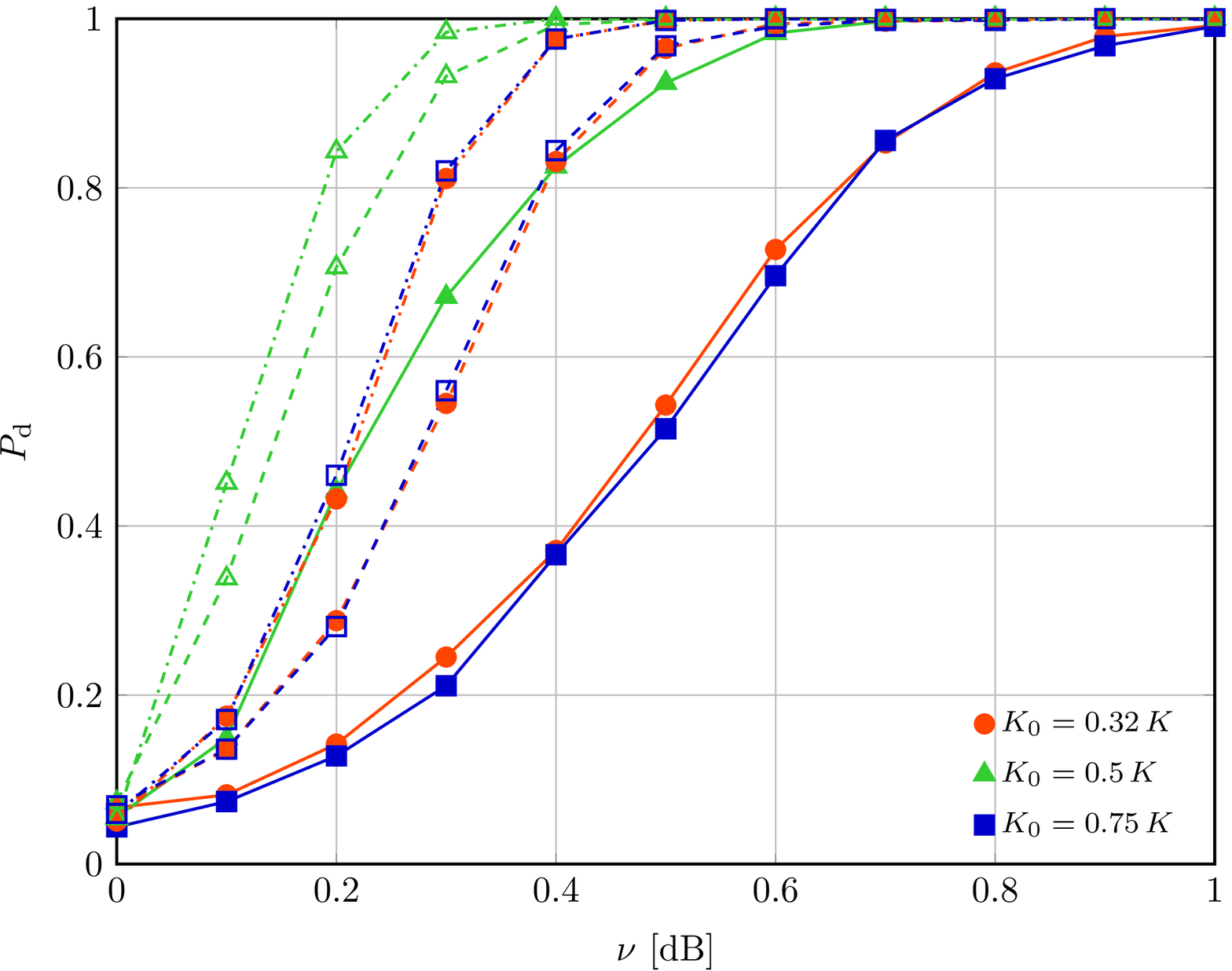}}
\end{subfigure}
\caption{$P_\text{d}$ versus $\nu$ for the SP-D-UM, SP-D-CM, and LVM-SP-D assuming $K=32$, \ac{SNR}$=-30\,$dB (solid lines), \ac{SNR}$=-15\,$dB (dashed lines), and \ac{SNR}$=0\,$dB (dot-dashed lines).}
\label{fig:Spoof32}
\end{figure*}

Starting from the case $K=24$, we notice that, unlike the performance of the NLJ detectors, the \ac{SNR} variations significantly 
affects the $P_\text{d}$ that, in this case, increases with the \ac{SNR}. This behavior can be observed for all the proposed spoofing detectors.
For instance, in Figure \ref{fig:SpoofUNC24}, given $\nu=0.2\,$dB and $K_0=K/2$,  $P_\text{d}=0.17$ for SNR$=-30\,$dB, $P_\text{d}=0.61$ for 
SNR$=-15\,$dB, and $P_\text{d}=0.94$ for SNR$=0\,$dB. Analogous remarks hold also for Figures \ref{fig:SpoofCOR24} and \ref{fig:SpoofLVM24}.
As for the loss of SP-D-CM with respect to the SP-D-UM due to the more general design assumptions 
of the former with respect to the latter, in this case, it is less important than that observed in the numerical examples of the 
previous subsection.
Moreover, notice that, differently from the LVM-NLJ-D, the $P_\text{d}$ curves associated with the LVM-SP-D achieve a value greater
than $0.9$ for each considered value of $K_0$.
Finally, in Figure \ref{fig:Spoof32} 
we evaluate the effect of $K$ on the detection performance that, again, 
improves as $K$ increases. Another
important aspect to be underlined is that for high $K$ values, the $P_\text{d}$ variation 
induced by $K_0$ reduces.

Summarizing, also in the presence of a spoofing attacks, we select the SP-D-CM as the suggested solution against this kind of attacks since
it accounts for a more general structure for the measurement covariance matrix and the loss with respect to the SP-D-UM becomes
negligible for high values of $K$.

\section{Conclusions}
\label{sec:conclusions}
In this paper we have addressed the problem of detecting an attack to the location services provided by the next generation 
communication networks. The proposed strategies have been conceived for high-level location data in agreement with 5G standardization work. %a feasibility that is as
%general as possible regardless the specific underlying network architecture. %As a matter of fact, the standardization
%of the location technologies that will be used in 5G networks is still in progress leaving the door open to different solutions.
%The proposed algorithms are fed by high-level location data in agreement with 5G standardization rather than raw (physical) data 
%representing a flexible approach that can be incorporated without additional hardware resources or architecture redefinition. 
At the design stage, this problem has been formulated as a binary hypothesis
test accounting for possible correlation among the measurements and solved resorting to GLRT-based design procedures as well as the \ac{LVM}.
The analysis conducted on simulated data has singled out the NLJ-D-CM and the SP-D-CM as an effective means to cope with NLJ and spoofing
attacks since they represent a good compromise between detection performance and capability of operating without any assumption on
data correlation.

Future research tracks might include the development of an architecture capable of classifying and detecting which kind of attack
is in course also accounting for more sophisticated jamming attackers.

%%%%%\section*{Acknowledgment}
%%%%%The authors wish to thank XYZ and ABC for their helpful suggestions
%%%%%and careful reading of the manuscript.

\bibliographystyle{IEEEtran}
\bibliography{IEEEabrv,MyBib,IvanBib}

\newpage
%---------------------------------------------------------------------------%
%                                   cover sheet                             %
%---------------------------------------------------------------------------%
{
\onecolumn % force the text after this point to be 1 column

\title{\paperTitle}
\author{Wgroup\\
		\vspace{0.5cm}
		\textcolor{BLUE}\versionDT
	   }
\maketitle

%\tableofcontents

%% ------ Catechism  ------ %%

%\newpage
%
%%\input{WINS-Catechism} 
%
%\bigskip
%Helpful hints:
%
%\vfill
%\noindent \textit{Template version \today}
}

\end{document}

%% file: LatexInclusion/Wgroup-Teaching-Preamble.tex
\usepackage{bm}

\DeclareMathAlphabet{\mathsfbr}{OT1}{cmss}{m}{n}%for math sans serif (cmss)
\SetMathAlphabet{\mathsfbr}{bold}{OT1}{cmss}{bx}{n}%for math sans serif (cmss)
\DeclareRobustCommand{\msf}[1]{%
  \ifcat\noexpand#1\relax\msfgreek{#1}\else\mathsfbr{#1}\fi%for math sans serif (cmss)
}
\makeatletter
\newcommand{\msfgreek}[1]{\csname s\expandafter\@gobble\string#1\endcsname}
\makeatother

\DeclareFontEncoding{LGR}{}{} % or load \usepackage{textgreek}
\DeclareSymbolFont{sfgreek}{LGR}{cmss}{m}{n}
\SetSymbolFont{sfgreek}{bold}{LGR}{cmss}{bx}{n}
\DeclareMathSymbol{\salpha}{\mathord}{sfgreek}{`a}
\DeclareMathSymbol{\sbeta}{\mathord}{sfgreek}{`b}
\DeclareMathSymbol{\sgamma}{\mathord}{sfgreek}{`g}
\DeclareMathSymbol{\sdelta}{\mathord}{sfgreek}{`d}
\DeclareMathSymbol{\sepsilon}{\mathord}{sfgreek}{`e}
\DeclareMathSymbol{\szeta}{\mathord}{sfgreek}{`z}
\DeclareMathSymbol{\seta}{\mathord}{sfgreek}{`h}
\DeclareMathSymbol{\stheta}{\mathord}{sfgreek}{`j}
\DeclareMathSymbol{\siota}{\mathord}{sfgreek}{`i}
\DeclareMathSymbol{\skappa}{\mathord}{sfgreek}{`k}
\DeclareMathSymbol{\slambda}{\mathord}{sfgreek}{`l}
\DeclareMathSymbol{\smu}{\mathord}{sfgreek}{`m}
\DeclareMathSymbol{\snu}{\mathord}{sfgreek}{`n}
\DeclareMathSymbol{\sxi}{\mathord}{sfgreek}{`x}
\DeclareMathSymbol{\somicron}{\mathord}{sfgreek}{`o}
\DeclareMathSymbol{\spi}{\mathord}{sfgreek}{`p}
\DeclareMathSymbol{\srho}{\mathord}{sfgreek}{`r}
\DeclareMathSymbol{\ssigma}{\mathord}{sfgreek}{`s}
\DeclareMathSymbol{\stau}{\mathord}{sfgreek}{`t}
\DeclareMathSymbol{\supsilon}{\mathord}{sfgreek}{`u}
\DeclareMathSymbol{\sphi}{\mathord}{sfgreek}{`f}
\DeclareMathSymbol{\schi}{\mathord}{sfgreek}{`q}
\DeclareMathSymbol{\spsi}{\mathord}{sfgreek}{`y}
\DeclareMathSymbol{\somega}{\mathord}{sfgreek}{`w}

\DeclareMathSymbol{\svarsigma}{\mathord}{sfgreek}{`c}

\DeclareMathSymbol{\sGamma}{\mathalpha}{sfgreek}{`G}
\DeclareMathSymbol{\sDelta}{\mathalpha}{sfgreek}{`D}
\DeclareMathSymbol{\sTheta}{\mathalpha}{sfgreek}{`J}
\DeclareMathSymbol{\sLambda}{\mathalpha}{sfgreek}{`L}
\DeclareMathSymbol{\sXi}{\mathalpha}{sfgreek}{`X}
\DeclareMathSymbol{\sPi}{\mathalpha}{sfgreek}{`P}
\DeclareMathSymbol{\sSigma}{\mathalpha}{sfgreek}{`S}
\DeclareMathSymbol{\sUpsilon}{\mathalpha}{sfgreek}{`U}
\DeclareMathSymbol{\sPhi}{\mathalpha}{sfgreek}{`F}
\DeclareMathSymbol{\sPsi}{\mathalpha}{sfgreek}{`Y}
\DeclareMathSymbol{\sOmega}{\mathalpha}{sfgreek}{`W}

\DeclareRobustCommand{\mcal}[1]{%
  \ifcat\noexpand#1\relax\mathnormal{#1}\else\cal{#1}\fi
}
\DeclareRobustCommand{\BM}[1]{%
  \ifcat\noexpand#1\relax\bm{\boldUppercaseItalicGreek{#1}}\else\bm{#1}\fi
}
\makeatletter
\newcommand{\boldUppercaseItalicGreek}[1]{\csname var\expandafter\@gobble\string#1\endcsname}
\makeatother
\newcommand{\V}[1]{\bm{#1}}
\newcommand{\M}[1]{\BM{#1}}

%% file: LatexInclusion/defmetric.tex
\DeclareMathAlphabet{\foo}{U}{tx-cal}{m}{n}

\def\R{{\mathds R}}
\newcommand{\tr}{\mbox{\rm Tr}\, }
\def\cN{\mbox{$\CMcal N$}}
\newcommand{\be}{\begin{equation}}
\newcommand{\ee}{\end{equation}}
\newcommand{\ds}{\displaystyle}
\newcommand{\diag}{\mbox{\rm \bf diag}\, }
\newcommand{\test}{\mbox{$
\begin{array}{c}
\stackrel{ \stackrel{\textstyle H_{1}}{\textstyle >} }{
\stackrel{\textstyle <}{\textstyle H_0} }
\end{array}
$}}
\newcommand{\dmax}{\begin{displaystyle}\max\end{displaystyle}}

\def\cA{{\mathcal A}}

\def\cL{\mbox{$\mathcal L$}}
\def\cN{\mbox{$\CMcal N$}}

\newcommand{\bd}{\begin{description}}
\newcommand{\ed}{\end{description}}
\newcommand{\ben}{\begin{enumerate}}
\newcommand{\een}{\end{enumerate}}
\newcommand{\bi}{\begin{itemize}}
\newcommand{\ei}{\end{itemize}}
\newcommand{\bl}{\begin{list}}
\newcommand{\el}{\end{list}}
\newcommand{\bt}{\begin{tabbing}}
\newcommand{\et}{\end{tabbing}}

\definecolor{BLUE}{rgb}{0,0,1}

\newcommand{\hsa}[1] {h_{\mathrm{s},\mathrm{a}}}

\newcommand{\srxb}[1] {r_{\mathrm{s-b},i}(t)}
\newcommand{\srxref}[1]{r_{\mathrm{ref},i}(t)}
\newcommand{\srxrem}[1]{r_{\mathrm{rem},i}(t)}

\newcommand{\thetaBi}[1]{\boldsymbol \theta_{B_i}}

\newcommand{\toar}[1]{\hat{\tau}_{i,0}}

\definecolor{myred}{rgb}{1,0.27,0}
\definecolor{mygreen}{rgb}{0.20, 0.8, 0.2}
\definecolor{myblue}{rgb}{0, 0, 0.8}
\definecolor{myorange}{RGB}{255, 178, 102}
\definecolor{mymagenta}{rgb}{0.78, 0.08, 0.52}
\definecolor{mycyan}{rgb}{0, 0.74, 1}

%% file: LatexInclusion/acronym.tex
\acrodef{ACK}{acknowledge}
\acrodef{AE}{angle estimate}
\acrodef{ADM}{angle direction matrix}
\acrodef{AI}{angle information}
\acrodef{AII}{angle information intensity}
\acrodef{AL}{angle likelihood}
\acrodef{AN}{Access Node}
\acrodef{AOA}{angle-of-arrival}
\acrodef{AWGN}{additive white Gaussian noise}
\acrodef{BC}{belief condensation}
\acrodef{BCF}{belief condensation filter}
\acrodef{BP}{belief propagation}
\acrodef{BPZF}{band-pass zonal filter}
\acrodef{BS}{base station}
\acrodef{BTB}{Bellini-Tartara bound}
\acrodef{CCDF}{complementary cumulative distribution function}
\acrodef{CDF}{cumulative distribution function}
\acrodef{CEO}{counting error outage}
\acrodef{CF}{characteristic function}
\acrodef{CFAR}{constant false alarm rate}
\acrodef{CIR}{channel impulse response}
\acrodef{CR}{channel response}
\acrodef{CRB}{Cram\'{e}r--Rao bound}
\acrodef{CRLB}{Cram\'{e}r--Rao lower bound}
\acrodef{CS}{case study}
\acrodef{DE}{distance estimate}
\acrodef{DFL}{device-free localization}
\acrodef{DL}{downlink}
\acrodef{DM}{direction matrix}
\acrodef{DOA}{direction of arrival}
\acrodef{DP}{direct positioning}
\acrodef{DRT}{distance ratio test}
\acrodef{DS}{delay spread}
\acrodef{ED}{energy detector}
\acrodef{EKF}{extended Kalman filter}
\acrodef{EFI}{equivalent Fisher information}
\acrodef{EFIM}{equivalent Fisher information matrix}
\acrodef{EM}{expectation-maximization}
\acrodef{EPF}{energy profile-based Particle filter}
\acrodef{ESD}{energy-based soft-decision}
\acrodef{ESPRIT}{estimation of signal parameters via rotational invariant techniques}
\acrodef{FCC}{Federal Communications Commission}
\acrodef{FDOA}{frequency difference of arrival}
\acrodef{FG}{factor graph}
\acrodef{FII}{Fisher information inequality}
\acrodef{FIM}{Fisher information matrix}
\acrodef{FL}{feature likelihood}
\acrodef{FP}{feature potential}
\acrodef{FW}{Fisher-Wald}
\acrodef{GDOP}{geometric dilution of precision}
\acrodef{GLRT}{generalized likelihood ratio test}
\acrodef{GPS}{Global Positioning System}
\acrodef{HDSA}{high-definition situation-aware}
\acrodef{HI}{hard information}
\acrodef{HMM}{hidden Markov model}
\acrodef{IID}{Independent and Identically Distributed}
\acrodef{IMU}{inertial measurement unit}
\acrodef{INR}{interference-to-noise ratio}
\acrodef{IoT}{Internet-of-Things}
\acrodef{IR-UWB}{impulse radio UWB}
\acrodef{JBSF}{jump back and search forward}
\acrodef{KDE}{kernel density estimation}
\acrodef{KF}{Kalman filter}
\acrodef{LBP}{loopy belief propagation}
\acrodef{LEM}{Laplacian eigen-map}
\acrodef{LEO}{localization error outage}
\acrodef{LOS}{line-of-sight}
\acrodef{LoT}{Localization-of-Things}
\acrodef{LRT}{likelihood ratio test}
\acrodef{LS}{least squares}
\acrodef{LTE}{long-term evolution}
\acrodef{LVM}{latent variable model}
\acrodef{MAC}{medium access control}
\acrodef{MAP}{maximum {\it a posteriori}}
\acrodef{MBS}{maximum bin search}
\acrodef{MDD}{minimum distance distribution}
\acrodef{MF}{matched filter}
\acrodef{MIMO}{multiple-input multiple-output}
\acrodef{ML}{maximum likelihood}
\acrodef{MLES}{Maximum Likelihood Estimation}
\acrodef{MLE}{Maximum Likelihood Estimate}
\acrodef{MMSE}{minimum-mean-square-error}
\acrodef{MOC}{multi-target overlap coefficient}
\acrodef{MP}{map potential}
\acrodef{MPC}{multi-path component}
\acrodef{MSE}{mean-square-error}
\acrodef{MUI}{multi-user interference}
\acrodef{MUSIC}{multiple signal classification}
\acrodef{NBI}{narrowband interference}
\acrodef{NLJ}{noise-like jammer}
\acrodef{NLJs}{noise-like jammers}
\acrodef{NLN}{network localization and navigation}
\acrodef{NLOS}{non-line-of-sight}
\acrodef{NR}{New Radio}
\acrodef{OFDM}{orthogonal frequency division multiplexing}
\acrodef{OP}{outage probability}
\acrodef{OTDOA}{observed time difference of arrival}
\acrodef{P-Max}{$P$-Max}  %suggestion, use with \acl{P-Max}
\acrodef{PAR}{probabilistic association rule}
\acrodef{PCA}{principal component analysis}
\acrodef{PDF}{probability density function}
\acrodef{PDP}{power delay profile}
\acrodef{PEB}{position error bound}
\acrodef{PF}{physical features}
\acrodef{PMF}{probability mass function}
\acrodef{POC}{path overlap coefficient}
\acrodef{PPM}{pulse position modulation}
\acrodef{PPP}{Poisson point process}
\acrodef{RCS}{radar cross section}
\acrodef{RDM}{range direction matrix}
\acrodef{RF}{radiofrequency}
\acrodef{RFID}{radio frequency identification}
\acrodef{RI}{range information}
\acrodef{RII}{range information intensity}
\acrodef{RL}{range likelihood}
\acrodef{RLS}{range-based least square}
\acrodef{RMS}{root mean square}
\acrodef{RMSE}{root-mean-square error}
\acrodef{RMU}{radio measurement unit}
\acrodef{RPF}{range-based Particle filter}
\acrodef{RRC}{root raised cosine}
\acrodef{RSS}{received signal strength}
\acrodef{RSRP}{reference signal received power}
\acrodef{RTT}{round-trip time}
\acrodef{RV}{random variable}
\acrodef{SBS}{serial backward search}
\acrodef{SBSMC}{serial backward search for multiple clusters}
\acrodef{SC}{soft constraint}
\acrodef{SDM}{speed direction matrix}
\acrodef{SK}{soft knowledge}
\acrodef{SFI}{soft feature information}
\acrodef{SCI}{soft context information}
\acrodef{SCNR}{signal to clutter plus noise ratio}
\acrodef{SI}{soft information}
\acrodef{SRI}{soft range information}
\acrodef{SAI}{soft angle information}
\acrodef{SII}{speed information intensity}
\acrodef{SIR}{signal-to-interference ratio}
\acrodef{SNR}{signal-to-noise ratio}
\acrodef{SO}{soft observation}
\acrodef{SPAWN}{sum-product algorithm over a wireless network}
\acrodef{mSPEB}{multi-target squared position error bound}
\acrodef{sSPEB}{single-target squared position error bound}
\acrodef{tSPEB}{total squared position error bound}
\acrodef{SPEB}{squared position error bound}
\acrodef{SR}{sensor radar}
\acrodef{SRN}{sensor radar network}
\acrodef{SVE}{single-value estimate}
\acrodef{TCS}{threshold crossing search}
\acrodef{TDL}{tapped delay line}
\acrodef{TDOA}{time difference of arrival}
\acrodef{TH}{time-hopping}
\acrodef{TNR}{threshold-to-noise ratio}
\acrodef{TOA}{time of arrival}
\acrodef{TOF}{time-of-flight}
\acrodef{TSD}{threshold-based soft-decision}
\acrodef{UE}{user equipment}
\acrodef{UKF}{unscented Kalman filter}
\acrodef{UMa}{urban macro}
\acrodef{UML}{unsupervised machine learning}
\acrodef{UWB}{ultrawide-band}
\acrodef{VDM}{velocity direction matrix}
\acrodef{WAF}{wall attenuation factor}
\acrodef{WBI}{wideband interference}
\acrodef{WED}{wall extra delay}
\acrodef{WLS}{weighted least squares}
\acrodef{WPAN}{wireless personal area network}
\acrodef{WSN}{wireless sensor network}
\acrodef{WWB}{Weiss-Weinstein bound}
\acrodef{ZZB}{Ziv-Zakai bound}
\acrodef{ZZLB}{Ziv-Zakai lower bound}